\documentclass[]{aa}  

\usepackage{graphicx}
\usepackage[varg]{txfonts}
\usepackage{hyperref}
\usepackage{xcolor}
\usepackage{amsfonts}
\begin{document}

   \title{Star-Planet Interaction:}

   \subtitle{Wave Structures and Wing-Wing Interaction}

   \author{C. Fischer
          \inst{1}
          \and
          J. Saur
          \inst{1}
          }

   \institute{Institute for Geophysics and Meteorology, University of Cologne,
              Pohligstra\ss e 3, 50969 Cologne\\
              \email{cfisch12@uni-koeln.de} \\
              \email{jsaur@uni-koeln.de}
             }

   \date{Received ; accepted }


  \abstract
   {Electromagnetic Star-Planet Interaction (SPI) describes the phenomenon, when a planet couples to its host star via electromagnetic forces. Alfvén waves can establish such a coupling by forming Alfvén wings. SPI allows phenomena that we do not know from the Solar System. Wing-wing interaction is such an example, where the Alfvén wings of two planets merge and interact non-linearly.}
   {In this paper we focus on the effects that SPI has on other planets and the stellar wind. First, we analyse the different wave structures connected to SPI. The second part then investigates wing-wing interaction.}
   {Our study applies a magnetohydrodynamic model to describe a stellar system with multiple possible planets. As an example, we chose TRAPPIST-1 and its two innermost planets. We extended the PLUTO code to simulate collisions between atmospheric neutral particles and plasma ions. Neutral gas clouds imitate the planets and move through the simulation domain. That allows the simulation of fully time-dependent stellar systems.}
   {We analysed the wave structures, which result from the interaction between stellar wind and TRAPPIST-1 b. The inward going wave structure is an Alfvén wing. The outward going part of the interaction consists of an Alfvén wing, slow mode waves, the planetary wake and a slow shock. We quantified the strength of the respective wave perturbations at the outer planets to be on the order of 10\% to 40\% of the local background values of thermal, magnetic and dynamic pressure. Wing-wing interaction occurs due to the relative position of two planets during their conjunction and shows three phases. First there is an initial, non-linear intensification of the Poynting flux by $20\%$, an intermediate phase with reduced Poynting flux and a third phase when the Alfvén wing of planet c goes through planet b's wave structures with another intensification of the Poynting flux.}
   {}
   

   \maketitle
%


\section{Introduction}
The electromagnetic coupling between a planet and its host star is commonly referred to as magnetic Star-Planet Interaction and in the following abbreviated as SPI. In that situation, the star creates a magnetised plasma environment for the planet through the stellar wind and magnetic field \citep{Strugarekea15,Saur18}. On its orbit, the planet moves through the plasma and magnetic field and excites different plasma waves. If the motion is partially perpendicular to the magnetic field, the planet excites Alfvén waves, which are the most important waves in SPI \citep{Neubauer80,Saurea13}. In the case that the planet orbits sufficiently close to its star that the wave speed exceeds the local plasma speed then the waves can travel upstream towards the star and establish a coupling \citep{Strugarekea19,Saurea13,Saur18}. This situation is commonly referred to as sub-Alfvénic interaction \citep{Neubauer80}. The principal process is well known from the moon-magnetosphere interaction in our Solar System.

The most common approaches to observe SPI are spectral observations of typical emission lines for stellar chromospheres and coronae. The earliest approach to observe SPI dates back to the early 2000's and focused on non-thermal chromospheric Ca II H and K emissions \citep{Shkolnikea03}. Subsequent studies by \cite{Shkolnikea05,Shkolnikea08} of the system HD 179949 saw enhanced emissions in four of six observational epochs that appeared to be periodic with the planetary orbital period. Other authors observed different systems and likewise saw chromospheric excess emissions \citep{Walkerea08,Staabea17}. \cite{Cauleyea19} estimated possible magnetic field strengths of Hot Jupiters based on chromospheric excess emissions. The derived energy fluxes from chromospheric Ca II H and K emissions are larger than typical fluxes derived from Alfvén wing models and magnetohydrostatic models \citep{Saurea13,Lanza15} and the observed periodicity only appears in some epochs, which was suggested to be alternatively explained by star spots \citep{Millerea12}. Recently, \cite{Strugarekea19} investigated how modelling and observations could reconcile by including the magnetic topology of stellar coronae based on the example of Kepler-78.
In the UV range, \cite{Franceea16,Franceea18} conducted large surveys with the Hubble Space Telescope but the authors could not identify signals related to SPI. 
At coronal Xray wavelengths, several studies investigated the influence of planets on stellar Xray-activity, e.g. \cite{Kashyapea08,Scharf10,Poppenhaegerea10}. Some studies found correlations between planets and Xray-activity in stars and some did not. \cite{PoppenhaegerSchmitt11,PoppenhaegerSchmitt11b} explain previously observed correlations, attributed to SPI, as a result of selection effects due to planet detection methods and the limitations in Xray-observations.
The radio wavelength range is particularly promising to search for SPI. Recent observations with LOFAR showed strong hints for the existence of SPI in several systems \citep{Vedanthamea20,Callinghamea21,PerezTorresea21}. Also recently, a campaign in the radio range provided the first tentative direct observational hints of an intrinsic magnetic field on an exoplanet \citep{Turnerea21}.

Apart from spectroscopic observations, there are alternative approaches that aim to identify SPI. One way is based on flare triggering through SPI as derived by \cite{Lanza18}. \cite{FischerSaur19} analysed the flares of TRAPPIST-1, which were observed by K2. In the light curve, the authors found tentative evidence for flares, triggered by TRAPPIST-1 c. \cite{HowardLaw21} performed a statistical study on a larger set of TESS flare data but could not clearly identify SPI.
Another way to observe SPI focuses on auroras on brown dwarfs. Planets in orbit around very faint objects like brown dwarfs could generate a plasma environment like Io does in Jupiter's magnetosphere. That would cause bright auroral ovals and auroral spots caused by SPI \citep{Saurea21}. Circular polarized radio emissions from brown dwarfs were interpreted as auroral emissions \citep{Hallinanea15,Kaoea16}.  Subsequent follow up observation in the UV with HST could not confirm auroral emission in the UV in the case of LSR J1835+3259 \citep{Saurea18}, but found tentative hints for auroral UV emission on 2MASS J1237+6526 \citep{Saurea21}.

Despite the advances in the observation of SPI, the research of SPI still also relies on modelling. The theoretical basis are the analytical models of the interaction between Jupiter and its moon Io. The unipolar inductor model by \cite{GoldreichLyndenBell69} is a first theoretical model that works in a tenuous plasma environment. Later, \cite{Goertz80} and \cite{Neubauer80} derived interaction models based on Alfvén waves, where the moon forms so-called Alfvén wings. The model by \cite{Neubauer80} became the standard model for moon-magnetosphere interactions. The model works for mechanical interactions as well as electromagnetic interactions \citep{Neubauer98}. \cite{Saurea13} then applied the Alfvén wing model to SPI. Alternative approaches to SPI that base purely on electromagnetic interaction have been proposed by \citet{Lanza09,Lanza12,Lanza13}. These models require a magnetized planet that couples to the coronal magnetic field of its host star. \citet{Lanza09} and \cite{Lanza12} propose an electric current system that closes through reconnection at the planetary magnetopause. Another model by \citet{Lanza13} bases on the accumulation and release of stresses in magnetic flux tubes. The two models assume magnetohydrostatic force-free magnetic fields that couple the stellar corona and a magnetized planet.

Apart from the analytic model branch, SPI has been investigated in a variety of magnetohydrodynamic (MHD) simulations. First numerical studies focused on the Hot Jupiters. These studies investigated, which planets may experience sub-Alfvénic conditions and determined Alfvén characteristics and related current systems \citep{Ipea04,Preusseea06,Preusseea07}. \citet{Koppea11} built up on observations of HD 179949 and $\upsilon$ And \citep{Shkolnikea03} and successfully modelled the proposed phase shift between planet and observed emission features. Simulations by \cite{Strugarekea12,Strugarekea14,Strugarekea16} investigated the interaction between magnetised and non-magnetised planets, each without an atmosphere, and the stellar wind plasma, to estimate the amount of electromagnetic angular momentum transfer between planet and star. The authors derived scaling laws for the involved torques that may be applied to explain planetary migration. \cite{Cohenea15} simulated the interaction between the atmosphere of a terrestrial planet and the stellar wind of an M-dwarf. A study carried out by \citet{Strugarekea15} investigated different stellar magnetic field configurations and determined the Poynting fluxes carried by the respective planetary Alfvén wing. Their findings are in accordance with the theory provided by \citet{Saurea13}, which supports the application of the Alfvén wing model in the context of SPI. \cite{DaleyYatesStevens18} determined the strength of planetary radio emissions emanating from the magnetospheres of Hot Jupiters with MHD simulations.   

Modelling time-variable stellar systems provides further insight into the processes connected to SPI but the simulation is computationally expensive and technically demanding. Principally, there are two options to achieve this goal. One is the fully time-variable MHD simulation with, e.g. moving planets inside the domain. The other option is computationally less expensive and combines MHD simulations with analytical modelling. The first fully time-dependent simulation of a planet orbiting a star was published by \citet{Cohenea11}. The authors applied an internal boundary for the planet and let the code update the position at each time step. Physically, the planet interacted with the stellar wind by an intrinsic magnetic field. The simulation showed that SPI could generate hot spots in stellar atmospheres. An example of the second approach is \cite{Strugarekea19}, who investigated the system Kepler-78. The authors applied observed magnetic field maps of the star and simulated the respective stellar wind. The expected time-variable SPI has been modelled analytically, based on the result of the stellar wind.

SPI relates to many aspects of space physics of exoplanets. In one way, SPI may cause an increased stellar activity through the power that Alfvén wings transport to the star. The flare triggering mechanism by \cite{Lanza18} is such an example. Flares would have implications for the planets and possible habitability. However, the space environment of exoplanets could also be affected by other planets. SPI consists of many different waves, which may alter the shape of upper atmospheric layers, influence the atmospheric mass loss, change the size and shape of magnetospheres and could even affect the auroral activity of planets. The open field line structure of stellar magnetic fields outside a certain source region \citep{Jardineea02a} allows for electromagnetic interactions between planets. In these cases, the planets which generate SPI form wave structures that point inward, i.e. towards the star, and outward, i.e. away from the star. In multi-planet systems the outward going waves will affect the stellar wind environments of outer planets. In the case that two (or more) planets generate SPI, there can be wave-wave interactions. \cite{FischerSaur19} call the phenomenon where the Alfvén wings of two planets merge and interact with each other 'wing-wing interaction'. This interaction is not possible in moon-magnetosphere interactions in the solar system because the giant planets have closed and nearly dipolar magnetic fields. This prohibits two moons to share the same field line. Wing-wing interaction will cause a variability in the power arriving at the star. The variability will appear periodically with the synodic period between both involved planets \citep{FischerSaur19}. Therefore, wing-wing interaction is a qualitatively new phenomenon that can only occur in SPI. Due to the expected variability of the energy flux that arrives at the star, wing-wing interaction is a potentially observable phenomenon, which is expected to reoccur periodically.

In this paper, we will investigate the different wave modes that are associated to SPI and then focus on wing-wing interactions. In sections \ref{section:SubAlfvenicInteraction} and \ref{section:Methods}, we will outline the required theoretical basis and the applied MHD model. The first part of our results in section \ref{section:WaveStructures} investigates the wave structures that appear in SPI. The second part of our results, in section \ref{section:WingWing}, presents the first simulation of SPI originating from two planets that undergo wing-wing interaction. Finally, we discuss the implications of our studies in section \ref{section:Conclusions}.

\section{Theory of Star-Planet Interaction \label{section:SubAlfvenicInteraction}}
In this section, we will present relevant theory that we apply to interpret our numerical results physically. To identify the occurring wave structures, we have to know how we can identify the respective waves. Therefore, we introduce linear MHD waves in section \ref{subsection: MHDwaves}. In section \ref{subsection:AWTheory}, we introduce the Alfvén wing model that we apply to interpret the occurring Alfvénic wave structures. Finally in section \ref{subsection:WaveWaveInteractions}, we look at the interaction of Alfvén waves with each other, which is a vital aspect of wing-wing interaction.

\subsection{MHD waves \label{subsection: MHDwaves}} 
MHD allows the existence of four, highly anisotropic wave modes. Through linearization and a plane wave approach, one can derive  dispersion relations and linear wave properties for the Alfvén Mode, the slow mode and the Fast Mode.

The Alfvén wave is a transversal, i.e. non-compressional, wave that bases on magnetic tension. It propagates parallel to the magnetic field with the so-called Alfvén speed
\begin{equation}
v_\mathrm{A} = \frac{B_0}{\sqrt{\mu_0 \rho_0}} \label{equation:AlfvenSpeed}
\end{equation}
with the background magnetic field strength $B_0$ and the mass density $\rho_0$. Alfvén waves generate perturbations in the velocity and the magnetic field, that are parallel or anti-parallel to each other, depending on the direction of propagation along the field line, and perpendicular to the background magnetic field. We will introduce the non-linear properties of this wave in section \ref{subsection:AWTheory} and will thus not discuss the linear wave properties here.

The slow mode, just like the Alfvén Mode, propagates parallel to the magnetic field. In contrast to the Alfvén Mode, this wave is a compressional, magnetosonic wave. When the speed of sound is smaller than the Alfvén speed, the slow mode waves propagate with the speed of sound
\begin{equation}
c_\mathrm{s} = \sqrt{\frac{\Gamma p_0}{\rho_0}} \label{equation:SoundSpeed} .
\end{equation}
with the pressure $p_0$ and the polytropic index $\Gamma$. A condition that is fulfilled in the cases that we study here. The wave generates a disturbance in the plasma velocity component that is parallel to the background magnetic field $\delta v_\parallel$. From there, the wave generates the following perturbations in density and pressure:
\begin{align}
\delta p & = \frac{p_0 \Gamma}{c_\mathrm{s}} \delta v_\parallel \label{equation:SMp}\\
\delta \rho & = \frac{\rho}{c_\mathrm{s}} \delta v_\parallel . \label{equation:SMrho}
\end{align}

The Fast Mode propagates into all directions. Perpendicular to the background magnetic field, the wave propagates with the magnetosonic wave speed, which is the euclidean norm of Alfvén and sound speed. Parallel to the magnetic field, the wave propagates with Alfvén speed since this wave mode does not generate wing-like wave structures because its wave speed into all directions is larger than the speed of the surrounding plasma.

The fourth wave mode is the Entropy Mode, which does not propagate on its own. The wave only generates a perturbation in density and temperature \citep{JeffreyTaniuti64,Neubauer98,Saur21} and forms the wake of the planet.

\subsection{Sub-Alfvénic Interaction \label{subsection:AWTheory}} 
Planets are obstacles to the flow of a magnetized plasma. If the space plasma environment is sub-Alfvénic, i.e. the plasma speed is lower than the Alfvén velocity, the interaction is similar to the moon-magnetosphere interactions of the giant planets' moons in our solar system. The major difference between moon-magnetosphere interaction and SPI is the geometry of plasma flow and magnetic field to each other \citep{Zarka07,Saurea13}. Also, similarly to moons, planets may interact with the plasma via an intrinsic magnetic field or a neutral atmosphere. 

An atmosphere acts as a mechanical obstacle for the plasma through collisions between plasma ions and atmospheric neutral particles. This results in a deceleration of the plasma and can be parametrized by a collision frequency $\nu_\mathrm{in}$. The collisions also generate an electrical conductivity perpendicular to the magnetic field characterized here through a  Pedersen conductance $\Sigma_\mathrm{P}$.  Together with the Alfvén conductance $\Sigma_\mathrm{A}$, $\Sigma_\mathrm{P}$ defines the strength of interactions \citep{Saurea13} as
\begin{equation}
\bar{\alpha} = \frac{\Sigma_\mathrm{P}}{\Sigma_\mathrm{P}+\Sigma_\mathrm{A}}. \label{equation:InteractionStrength}
\end{equation}

The interaction strength $\bar{\alpha}$ is a parameter that assumes values between $0$ and $1$, where $\bar{\alpha}=0$ implies that there is no interaction and $\bar{\alpha}=1$ implies a maximum deceleration of the plasma.

Through any of the two interaction types, the planet generates waves. It especially generates Alfvén waves that, under sub-Alfvénic conditions, form so-called Alfvén wings \citep{Neubauer80}. In the reference frame of the wave generating moon or planet, the Alfvén wing appears as a standing wave structure. From a frame of reference of plasma flow surrounding the moon, Alfvén wings are time-variable structures that move together with the planet. 

Alfvén wings follow a path that is defined by the Alfvén wave characteristics
\begin{equation}
   \textit{\textbf{c}}_\mathrm{A}^\pm = \textit{\textbf{v}}_\mathrm{rel} \pm \textit{\textbf{v}}_\mathrm{A} \label{eq:Alfvencharacteristic} .
\end{equation}
In that case $\textit{\textbf{v}}_\mathrm{rel}$ is the relative velocity between the plasma flow and the planet and $\textit{\textbf{v}}_\mathrm{A}$ is the local Alfvén velocity. According to \cite{Neubauer80}, the Alfvén characteristics assume an angle $\Theta_\mathrm{A}$ to the background magnetic field that is defined by  
\begin{equation}
\sin \Theta_\mathrm{A}^\pm = \frac{ M_\mathrm{A} \cos \Theta }{ \sqrt{1 + M_\mathrm{A}^2 \pm 2 M_\mathrm{A} \sin \Theta} } \label{equation:AlfvenAngle} .
\end{equation}
In this equation $M_\mathrm{A}$ is the Alfvén Mach number and $\Theta$ is the angle between the plasma velocity and the normal to the magnetic field.

Finally, Alfvén waves assume the following wave properties:
\begin{align}
\textit{\textbf{B}}_\perp & = \mu_0 \Sigma_\mathrm{A} \frac{\textit{\textbf{c}}_\mathrm{A}^\pm }{|\textit{\textbf{c}}_\mathrm{A}^\pm|} \times \textit{\textbf{E}} \label{equation:AWnonlinBperp} \\
B_\parallel & = \pm \sqrt{B_0^2 - B_\perp^2} \label{equation:AWnonlinBparallel}
\end{align}
with the motional electric field $\textit{\textbf{E}} = - \textit{\textbf{v}} \times \textit{\textbf{B}}$. The perturbations are perpendicular to the background magnetic field and the wave generates no perturbation of the magnetic field strength. Equation \ref{equation:AWnonlinBperp} is the polarization relation of the Alfvén wave.

\subsection{Interactions between Alfvén Waves \label{subsection:WaveWaveInteractions}} 
In the Solar System, the direct interaction between two Alfvén wings can never occur. The reason is that Jupiter's magnetic field is still approximately dipolar at the location of its main moons, in constrast to the sun's magnetic field, which is dominated by the radial flow of the stellar wind sufficiently far away from the sun. Therefore, two moons will never share the same field line. In the open magnetic field lines of stars, however, two planets may be able to share the same magnetic field line. That allows wave-wave interactions in the form of wing-wing interaction, where the Alfvén wings of two planets merge.

The interaction between Alfvén waves is well studied in a large number of applications in space and astrophysical plasmas and it is for example also the basis for MHD turbulence \citep{Elsasser50,Iroshnikov63,Kraichnan65,SridharGoldreich94,GoldreichSridhar95}. The non-linear nature of the interaction of counter-propagating Alfvén waves has thereby been established in many theoretical \citep[e.g.][]{Elsasser50,GoldreichSridhar95,HowesNielson13} and numerical studies \citep[e.g.][]{MaronGoldreich01,Jacobsenea07} as well as in laboratory experiments \citep{Drakeea13,Howesea12}.

\section{Simulating Planetary Systems \label{section:Methods}}
In our simulation we chose a setup, where the star is in the center and the planet at a certain distance. That allows us to simulate a stellar system with more than one planet.

To perform such a simulation, we apply the PLUTO MHD Code \citep{Mignoneea07}. The code solves single-fluid MHD equations in conservation form:
\begin{align}
\dfrac{\partial \rho}{\partial t}+\nabla \cdot (\rho \textbf{\textit{v}}) & =0  \label{equation:PLUTOContiCons}\\
 \dfrac{\partial \textbf{\textit{m}}}{\partial t} + \nabla \cdot \left( \textbf{\textit{m}} \textbf{\textit{v}} - \textbf{\textit{B}}\textbf{\textit{B}} + (p + \frac{B^2}{2})\mathbb{I} \right) & = - \rho \nabla \Phi \label{equation:PLUTOEoMCons}\\
\dfrac{\partial}{\partial t}(E_\mathrm{tot} + \rho \Phi) + \nabla \cdot \left( (E_\mathrm{tot} + p + \rho \Phi)\textbf{\textit{v}} - \textbf{\textit{B}}(\textbf{\textit{v}} \cdot \textbf{\textit{B}}) \right) & = 0 \label{equation:PLUTOEoECons} \\
\dfrac{\partial \textbf{\textit{B}}}{\partial t} + \nabla \cdot ( \textbf{\textit{v}}\textbf{\textit{B}} - \textbf{\textit{B}}\textbf{\textit{v)}} & =  0. \label{equation:PLUTOInductionCons}
\end{align}

Note that the equations \ref{equation:PLUTOContiCons} to \ref{equation:PLUTOInductionCons} are written in normalized units as defined by \cite{Mignoneea07} and the PLUTO userguide. Equation \ref{equation:PLUTOContiCons} is the continuity equation, which describes mass conservation through the mass density $\rho$ and the plasma velocity $\textbf{\textit{v}}$. The equation of motion (Equation \ref{equation:PLUTOEoMCons}) describes the conservation of momentum with the momentum density $\textbf{\textit{m}}=\rho \textit{\textbf{v}}$ and holds responsible for the evolution of the plasma velocity $\textbf{\textit{v}}$. The momentum flux is affected by the thermal pressure $p$ and the magnetic field $\textbf{\textit{B}}$. Equation \ref{equation:PLUTOEoECons} describes the evolution of the total energy $E_\mathrm{tot} = \rho e + m^2/(2 \rho) + B^2/2$, consisting of the internal energy, kinetic energy and magnetic energy respectively, given in normalized units. Equation of motion and Equation of energy include the gravitational potential $\Phi$, which represents the stellar gravitational field. The stellar wind is described by a polytropic law with a polytropic index of $\Gamma = 1.2$. The chosen value is in accordance with simulations of the solar wind, which typically assume polytropic indices between 1 and about 1.2 \citep[e.g.][]{KeppensGoedbloed00,Roussevea03,Jacobsea05} and empirical estimates that provide values of up to 1.5 \citep{Tottenea95}. The evolution of the magnetic field is described by the induction equation \ref{equation:PLUTOInductionCons}. We apply the ideal MHD equations without viscosity and magnetic diffusion because the phenomena we are interested in, i.e. Alfvén wings, are not affected by diffusive effects on the spatial scales of this investigation. The regions of the stellar atmosphere that are heavily affected by diffusive effects, e.g. the stellar chromosphere and corona, are not directly present in our simulation.

We apply spherical coordinates and simulate on a grid that forms a spherical shell segment. The star is the center of our coordinate system. The boundary zone facing the star is referred to as the inner boundary with radius $R_\mathrm{inner}$, whereas the other end of the simulation domain is called the outer boundary. The inner boundary does not lie at the stellar surface, but instead it lies further away at the critical point of the Parker-model \citep{Parker58}. At this point, the stellar wind reaches the sound speed, giving super-sonic conditions further out. We set the radius of the inner boundary  to be slightly outside this point, i.e. $R_\mathrm{inner} = 5 R_*$ with the stellar radius $R_*$. The outer boundary has to reside outside the Alfvén radius of the simulated stellar wind. According to \cite{ZanniFerreira09}, this prevents Alfvén waves that come from an suboptimal outflow boundary and move inward towards the star. The outer boundary's radius lies at $R_\mathrm{outer} = 70 R_*$.

At the inner boundary, we pose Dirichlet-type inflow conditions for
\begin{align}
 v_r & = c_\mathrm{s}  \\
 \rho & = \rho_0 \left( \frac{R_*}{R_\mathrm{inner}} \right)^2 \frac{v_c}{v_r} 
\end{align}
and the pressure $p$ from the ideal equation of state with the coronal temperature $T$. The radial velocity $v_r$ is set at the speed of sound $c_\mathrm{s}$, the density follows a radial profile provided from the continuity equation with the plasma velocity estimated for the corona $v_c$.

For the magnetic field and the remaining velocity components, we apply outflow conditions that ensure there is no derivative normal to the boundary surface:
\begin{align}
\frac{\partial v_\mathrm{i}}{\partial n} & = 0 \\
\frac{\partial B_\mathrm{i}}{\partial n} & = 0
\end{align}
where the index $i$ represents the different components $r$, $\vartheta$, $\varphi$. The remaining five boundary zones apply the outflow conditions provided by PLUTO.

The stellar rotation needs to be specifically addressed by boundary conditions. Therefore, in the zone between $R_\mathrm{inner}$ and $R_\mathrm{inner} + 1 R_*$, we apply an internal boundary condition. The condition defined by \cite{ZanniFerreira09} for the azimuthal components of velocity $v_\varphi$ and magnetic field $B_\varphi$ ensures an accurate representation of the stellar rotation and a proper coupling of magnetic field and plasma in the boundary zone.

As initial conditions, we apply a previously simulated steady state stellar wind. To obtain this initial setup, we applied the stellar wind model described by \cite{FischerSaur19}, which is based on the Parker-model \citep{Parker58}.

\subsection{Modelling the Planets} 
In this section we will outline of how we model the interaction between planets and plasma. We saw in section \ref{subsection:AWTheory} that planets can interact with their plasma environment by collisions between atmospheric neutrals and plasma ions. In our simulations, we will apply this mechanism and implement the corresponding terms into the numerical model.

\subsubsection{Parametrisation of the Planet}
The planet is parametrized as a neutral gas cloud. This approach neglects the existence of a solid planetary body, for the case of terrestrial planets, and reduces the planet to be a pure generator of MHD waves. The chosen approach is well suited for studies of the far-field wave structures that we are interested in. Furthermore, it allows a first simple setup, where we can simulate moving planets that interact with a stellar wind.

The gas cloud is radially symmetric about the center of the planet. The easiest way to define the cloud is to switch into planetocentric coordinates, with the planetographic distance 
\begin{equation}
r_\mathrm{p} = | \textit{\textbf{a}} - \textit{\textbf{r}}_\mathrm{n} |  \label{equation:DistancetoPlanet} ,
\end{equation}
where $\textit{\textbf{a}}$ is the position of the planet and $\textit{\textbf{r}}_\mathrm{n}$ the position of a neutral gas parcel. Inside the planet, i.e. for $r_\mathrm{p} < R_\mathrm{p}$ with the planetary radius $R_\mathrm{p}$, we assume a constant neutral gas density of $n_0$. For the atmosphere, i.e. $r_\mathrm{p} > R_\mathrm{p}$, we assume a neutral gas density profile that follows a scale height law of
\begin{equation}
  n_\mathrm{n} = n_\mathrm{n,0} \exp \left( \frac{R_\mathrm{p} - r_\mathrm{p}}{H} \right) \label{equation:SimNeutralGas}
\end{equation}
where $H$ is the scale height of the atmosphere.

\subsubsection{Ion-Neutral Collisions in the Model Equations}
To simulate SPI via mechanical interactions, we have to expand PLUTO's model equations by source terms that model ion-neutral collisions. We apply the terms from \cite{Chaneea13}, which include the motion of the interacting body and are suitable for the conservative form of the MHD equations. For our model, the equations take the following form:
\begin{align}
\dfrac{\partial \textbf{\textit{m}}}{\partial t} + \nabla \cdot \textbf{\textit{F}}_\mathrm{m} & =  \mathrm{rhs}_\mathrm{m} -  \sum^n_{k=1} \rho \nu_\mathrm{in,k} ( \textbf{\textit{v}} - \textbf{\textit{v}}_\mathrm{orb,k}) \label{equation:SPIEoM}\\
\dfrac{\partial E_\mathrm{tot}}{\partial t}  + \nabla \cdot \textbf{\textit{F}}_\mathrm{E} & =  \\ 
\sum^n_{k=1} \nu_\mathrm{in,k} & \left(  - \frac{p}{\Gamma-1}   + \frac{1}{2} \rho ( \textbf{\textit{v}} - \textbf{\textit{v}}_\mathrm{orb,k} )^2 - \rho ( \textbf{\textit{v}} - \textbf{\textit{v}}_\mathrm{orb,k} ) \cdot \textbf{\textit{v}} \right) . \label{equation:SPIEoE} \nonumber
\end{align}
In this form, we can simulate a system with $n$ planets. Each of them will contribute separately to the collision terms via their individual collision frequency $\nu_\mathrm{in,k}$ and orbital velocity $\textbf{\textit{v}}_\mathrm{orb,k}$. Further variables are the momentum density $\textit{\textbf{m}}$, the total energy $E_\mathrm{tot}$, their respective fluxes $\textbf{\textit{F}}_\mathrm{m}$ and $\textbf{\textit{F}}_\mathrm{E}$ and the right hand side rhs$_m$ of equation \ref{equation:PLUTOEoMCons}. The chosen notation represents equations \ref{equation:PLUTOEoMCons} and \ref{equation:PLUTOEoECons} in shorter form.  The collision term in the equation of motion describes the exchange of momentum between plasma ions and atmospheric neutrals. This exchange is proportional to the relative velocity between plasma and neutral atmosphere. For the velocity of the neutral atmosphere, we only assume the orbital motion of the planet because it dominates over the speed caused by the planetary rotation and any kind of atmospheric dynamics. The terms in the energy equation describe the exchange of kinetic energy and internal energy.

We estimate the collision frequency $\nu_\mathrm{in}$ with the empirical formula

\begin{equation}
\nu_\mathrm{in} = 2.99 \cdot 10^{-15} n_\mathrm{n}(\textbf{\textit{r}}) \label{equation:collfreq}
\end{equation}

where the collision frequency $\nu_\mathrm{in}$ has units of $\mathrm{s}^{-1}$ and the neutral density $n_\mathrm{n}$ is provided in units of $\mathrm{m}^{-3}$. Expression \ref{equation:collfreq} is based on (9.73) from \cite{BanksKockarts73} with a polarisability of $\alpha_0 = 0.667 \cdot 10^{-30} \mathrm{m}^3$ and a reduced mass of $\mu_\mathrm{A} = 0.5038$ atomic mass units corresponding to proton-hydrogen collisions. The aeronomic processes in an exoplanet's atmosphere and ionosphere are much more complex than in our description. Details on such descriptions in sub-Alfvénic moon-planet interaction can for example be found in \cite{Neubauer98,Saurea98,Saurea99,Rubinea15} or \cite{Hartkornea17}. The purpose of this work is however not the detailed description of the immediate environment around planets with observationally constrained neutral atmospheres, but the far-field wave structure in sub-Alfvénic interaction.

We note that ion-neutral collisions within an atmosphere are one of several ways how planetary objects can generate sub-Alfvénic interactions \citep[see, e.g., review by][]{Saur21}. Intrinsic magnetic fields are the second main cause of sub-Alfvénic interactions (see section \ref{section:SubAlfvenicInteraction}). The third possibility is a rocky planetary body without a gaseous envelope or magnetic field of its own. In our study, the planets are primarily generators of waves and the aim is not to study the planets themselves. Therefore we choose ion-neutral collsions as the source of SPI.

\subsubsection{Motion of the Planet}
We aim to simulate a fully time-variable system with several planets. Hence, we have to include planetary motion into our model. We achieve this by making the collision frequency of the planetary atmosphere dependent on the planet's position. Under the assumption of circular orbits, the position is updated at every time step according to
\begin{equation}
	\textit{\textbf{a}}(t) = a
	\begin{pmatrix}
	\cos(\Omega_p t + \varphi_\mathrm{init}) \\
	
	\sin(\Omega_p t + \varphi_\mathrm{init}) \\
	
	0
	\end{pmatrix}.
\end{equation} 
We assume that at $t=0$ hr the planet is located at an azimuth $\varphi_\mathrm{init}$, possesses a colatitude of $90^\circ$ and is located at a distance $a$ from the star. Every time step, we rotate the position of the planet about the z-axis by the orbital angular velocity of the planet $\Omega_p$. In our case the z-axis coincides with the spin-axis of the star. The respective current position $\textbf{\textit{a}}$ then goes into the calculation of the collision frequency. In that manner, the collision frequency, as the important interaction parameter for the MHD equations, will be updated at every time step and simulates the motion of the planet around the star.

\subsection{Parameters} 
We will base our studies on the example of the TRAPPIST-1 system. TRAPPIST-1 is an M8 dwarf star with seven terrestrial planets in close orbits around the star \citep{Gillonea16,Gillonea17}. We know from previous studies that several of TRAPPIST-1's planets are likely candidates for the generation of SPI. \citet{Garraffoea17} performed MHD simulations of the system and estimated the Alfvén radius, i.e. the region wherein the Alfvén wave speed is larger than the plasma speed. The authors estimated that six of the seven planets potentially orbit within the Alfvén radius. A theoretical study by \citet{Turnpenneyea18} estimated all planets to be able to generate SPI. \citet{FischerSaur19} estimated that at least the two inner planets are likely to generate SPI. The authors performed a large parameter study with a theoretical stellar wind model. They found out that the most important parameter is the currently unknown stellar wind mass flux. This parameter finally defines, which of the planets will generate SPI.

We will apply the best guess model for the stellar wind from \cite{FischerSaur19}, which gives a coronal number density of $n_\mathrm{c} = 2.1 \cdot 10^{15} \, \mathrm{m}^{-3}$, a coronal temperature of $T_\mathrm{c} = 2 \, 000 \, 000 \, \mathrm{K}$ and a magnetic field strength at the stellar equator of $B_0 = 0.06 \, \mathrm{T}$. At the inner boundary $R_\mathrm{inner}$, these conditions produce a sound speed of $c_\mathrm{s} = 130 \, \mathrm{km}\mathrm{s}^{-1}$, which resembles the expected stellar wind speed. We further assume the rotation period of TRAPPIST-1 of $3.3$ days \citep{Lugerea17}.

The existence and thus the physical properties of the atmospheres of TRAPPIST-1 b and c are largely unknown. Early observations of TRAPPIST-1 b and c were conducted by \cite{DeWittea16} and \cite{Bourrierea17}. \cite{DeWittea16} ruled out cloud-free hydrogen dominated atmospheres, which implies that the planets either have no atmosphere or an atmosphere that consists of heavier molecules. \cite{Bourrierea17} estimated from Lyman-$\alpha$ observations of both planets conducted with HST that both planets might have extended hydrogen exospheres. \cite{Turbetea20} review the current state of knowledge for the planetary atmospheres in the TRAPPIST-1 system, including several physical properties of the system, observations of the planets and atmospheric modeling. The authors conclude that the most likely atmospheres for the inner planets of the system are either (1) water dominated, due to the strong irradiation and possible high water mass fraction of the planets, (2) $O_2$-dominated atmospheres, which are remnants of former water rich atmospheres or (3) $CO_2$-rich atmospheres, because this molecule is very resilient against atmospheric escape and also quite common in the solar system planets. Although, \cite{DeWittea16} ruled out cloud-free hydrogen dominated atmospheres at lower layers, we assume that the upper layers of planetary atmospheres typically consist of atomic hydrogen whereas the lower layers are usually dominated by different gas species. This assumption is supported by the observations of \cite{Bourrierea17}. Even if there are no gaseous envelopes present at TRAPPIST-1 b and c, the example of Saturn's moon Rhea shows that Alfvén wings can even form without a planetary atmosphere \citep{Simonea12,Khuranaea17}. In that case, the planet would absorb the plasma on its upstream side and generate a wake that is slowly filled with plasma. The resulting pressure gradient generates a differential plasma velocity, which causes magnetic perturbations that form an Alfvén wing.

We choose a base neutral gas density of $n_\mathrm{n,0} = 10^{14} \, \mathrm{m}^{-3}$, which produces a maximum interaction strength (see Equation \ref{equation:InteractionStrength}) of about 0.8 in the center of the planet. To resolve the atmosphere numerically and not only the planetary body, we apply a scale height of $H = 1300$ km. Due to the radially decreasing gravity in planetary atmospheres, the scale height can easily reach those values. 

\section{Wave Structures related to SPI \label{section:WaveStructures}}
In this section, we will analyse the different wave modes that are associated with SPI and investigate the influence of these waves on other planets. Our analysis is a basic study, which aims at a deeper understanding of SPI because the said wave modes can affect the space environments of other planets and may cause observable effects in the atmospheres and magnetospheres of planets. Detailed knowledge of the wave structures is especially required for the analysis of wing-wing interaction.

For the analysis of the planetary wave structures, we assume a single-planet scenario with TRAPPIST-1 and its closest planet TRAPPIST-1 b. The planet orbits the star at a distance of $r=20.5 \ R_*$. At the beginning of the simulation at $t=0$ s, the planet is located at $\vartheta = 90^\circ$, i.e. the equatorial plane, and $\varphi=14.3^\circ$. For the analysis, we choose the situation at $t = 5.7$ hr, when the planet is located at $\varphi=71^\circ$, i.e. $x = 6.6 R_*$ and $y = 19.3 R_*$.

Within the radial stellar wind and the open magnetic field line structure of stars the wave structures generated by planets have an inward pointing side and an outward pointing side. We will investigate both, inward and outward going  structures and identify the different waves.

\subsection{Wave pattern \label{section:WavePattern}} 

\begin{figure*}[t]
\centering
\includegraphics[width=17cm]{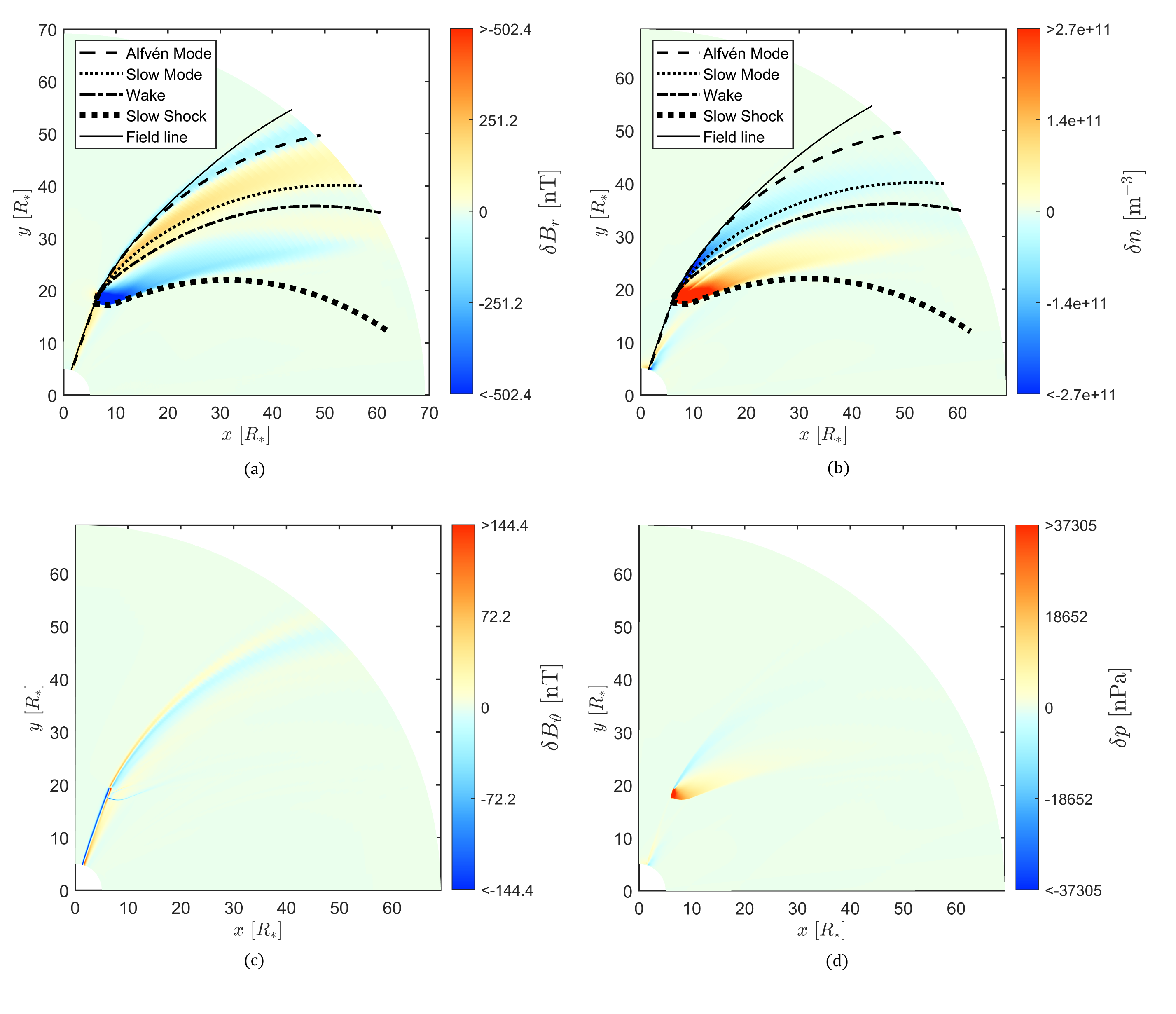}
\caption{Wave structures generated by TRAPPIST-1 b. Each panel shows the wave perturbations of a different plasma variable, where panels (a) and (b) also indicate the wave paths. Panel (a) shows the perturbation of the radial magnetic field component $\delta B_\mathrm{r}$, panel (b) the density perturbation $\delta \rho$, (c) the perturbation of the meridional magnetic field component $\delta B_\mathrm{\vartheta}$ and panel (d) shows the pressure perturbation $\delta p$.}
\label{fig:Characteristics}
\end{figure*}

In a first step, we will describe the different waves by means of their disturbances. Therefore, we show the equatorial planes of the simulation domain with the disturbances of chosen plasma quantities. Those are, the radial magnetic field disturbance $\delta B_\mathrm{r}$ (Figure \ref{fig:Characteristics} (a)), the density disturbance $\delta \rho$ (Figure \ref{fig:Characteristics} (b)), the meridional magnetic field disturbance $\delta B_\vartheta$ (Figure \ref{fig:Characteristics} (c)) and the pressure disturbance $\delta p$ (Figure \ref{fig:Characteristics} (d)). Blue colors represent negative disturbances and red colors positive disturbances. The colorbars of each plot are capped to values smaller than the absolute largest disturbances and centered around zero. The reason for this is to make the disturbances better visible throughout the simulation domain. The colour scale of $\delta B_r$ was reduced to 1/6 of the maximum value and to 1/2 for $\delta \rho$.

As a first quantitative analysis we describe the wave paths by

\begin{align}
\textbf{\textit{c}}_\mathrm{A}^\pm & = \textbf{\textit{v}} \pm \frac{\textbf{\textit{B}}}{\sqrt{\mu_0 \rho}} \label{equation:AlfvenCharacteristics}\\
\textbf{\textit{c}}_\mathrm{SM}^\pm & = \textbf{\textit{v}} \pm \frac{\textbf{\textit{B}}}{|\textbf{\textit{B}}|} c_\mathrm{s} \label{equation:SlowCharacteristics}\\
\textbf{\textit{c}}_\mathrm{w} & = \textbf{\textit{v}}. \label{equation:WakeCharacteristics3}
\end{align}

Equation \ref{equation:AlfvenCharacteristics} describes the Alfvén characteristics \citep{Elsasser50, Neubauer80}. Equations \ref{equation:SlowCharacteristics} and \ref{equation:WakeCharacteristics3} have been constructed from the respective wave properties. For the wave path of the slow mode, also called slow mode wing in the case of a standing wave, we used the fact that it assumes an angle $\Theta_\mathrm{s} = \tan^{-1} (M_\mathrm{s})$ to the background magnetic field, with the sonic Mach number $M_\mathrm{s} = v / c_\mathrm{s}$, and that the slow mode propagates along the magnetic field in case that $v_\mathrm{A} \gg c_\mathrm{s}$ \citep{WrightSchwartz90,Linkerea91}.

To estimate the path of each wave, we solve

\begin{equation}
 \frac{d \textit{\textbf{r}}}{d h} = \frac{\textit{\textbf{c}}_\mathrm{wave}}{|\textit{\textbf{c}}_\mathrm{wave}|}
\end{equation}

with expressions \ref{equation:AlfvenCharacteristics} to \ref{equation:WakeCharacteristics3} for $\textit{\textbf{c}}_\mathrm{wave}$ and the respective line element along the wave paths $dh$. With a Runge-Kutta Scheme, we can calculate the path of each wave. The starting point is the location of the planet.

We see in \ref{fig:Characteristics} (a) and (b) that the calculated wave paths follow certain structures that are visible in the magnetic disturbance and the density. The planet moves in anti-clockwise direction and the outward going waves follow a certain sequence according to their effective wave speed, i.e. the absolute value of the wave paths. The leading wave is the fastest wave, accordingly followed by the slower waves. However, the first structure is the magnetic field line (solid line) that connects to the planet.

The leading outward going wave mode, indicated by the dashed line, is the Alfvén Mode. The wave is visible in Figure \ref{fig:Characteristics} (a) and (c), i.e. the radial and meridional magnetic field perturbation. As we see in both plots, the Alfvén wave is the only wave mode that connects to the star.

At the position of the planet, the inward pointing wave assumes an angle of $2.6^\circ$ to the background magnetic field. This value fits well with the theoretically expected value of $\Theta_\mathrm{A} = 2.7^\circ$ according to equation \ref{equation:AlfvenAngle}.

Going further, in clockwise direction, there is a region of increased $B_r$ and decreased $\rho$ (Figure \ref{fig:Characteristics}) that coincides with the path of the outward going slow mode (dotted line). We see that the suspected wave generates negative disturbances in the density (Figure \ref{fig:Characteristics} (b)) and the pressure (Figure \ref{fig:Characteristics} (d)). The related wave structure broadens with distance from the planet. The group velocity of the slow mode can be non-parallel to the magnetic field \citep{Linkerea88}, just not perpendicular to the field. Therefore, with increasing distance from the planet, the wave widens up. 

Behind the outward going slow mode follows the planetary wake (dot-dashed line), which is the Entropy Mode. We see no apparent disturbances in the magnetic field or the pressure. Only the density shows a narrow increase that lies beneath the respective wave path (dot-dashed line), which strictly follows the flow of the plasma.

The last wave structure in clockwise direction poses an interesting case. Theoretically, the slowest outward going structure would be the wake because it does not propagate. Now, the existence of a disturbance in pressure and density indicates that there is a structure whose effective velocity is slower than the plasma flow and that cannot be explained by MHD wave theory. The structure moves a few stellar radii towards the star along the inward going Alfvén wing. Then it bends away from the Alfvén wing and moves outward, forming a broad compressional wave structure. We find that the respective structure fits well with the wave path of the inward going slow mode (thick dotted line), i.e. an effective wave speed of $v_\mathrm{eff} = v_r - c_\mathrm{s}$. Since we are in the super-sonic regime, one expects that the slow mode generates a Shock. Slow shocks propagate with a velocity of $v-c_\mathrm{s}$ \citep{Wuea96}, which supports our finding here. Apparently, the interaction of the planet generates an outward going slow mode wave that does not happen to be a shock because the waves never ramp up. The inward going slow mode, however, behaves differently. We will discuss the properties of both wave structures, that we will collectively call the Slow Structures, in section \ref{subsection:SloMoSloSho}.

The Fast Mode is the fourth wave mode in MHD. It is weakly anisotropic, but propagates into all directions, in contrast to the other MHD waves. The wave amplitude therefore declines rapidly with distance to the planet. Thus, we will not look further into the Fast Mode.

\subsection{Alfvén Wings \label{subsection:AlfvenWings}} 

\begin{figure*}[t]
\centering
\includegraphics[width=0.9\textwidth]{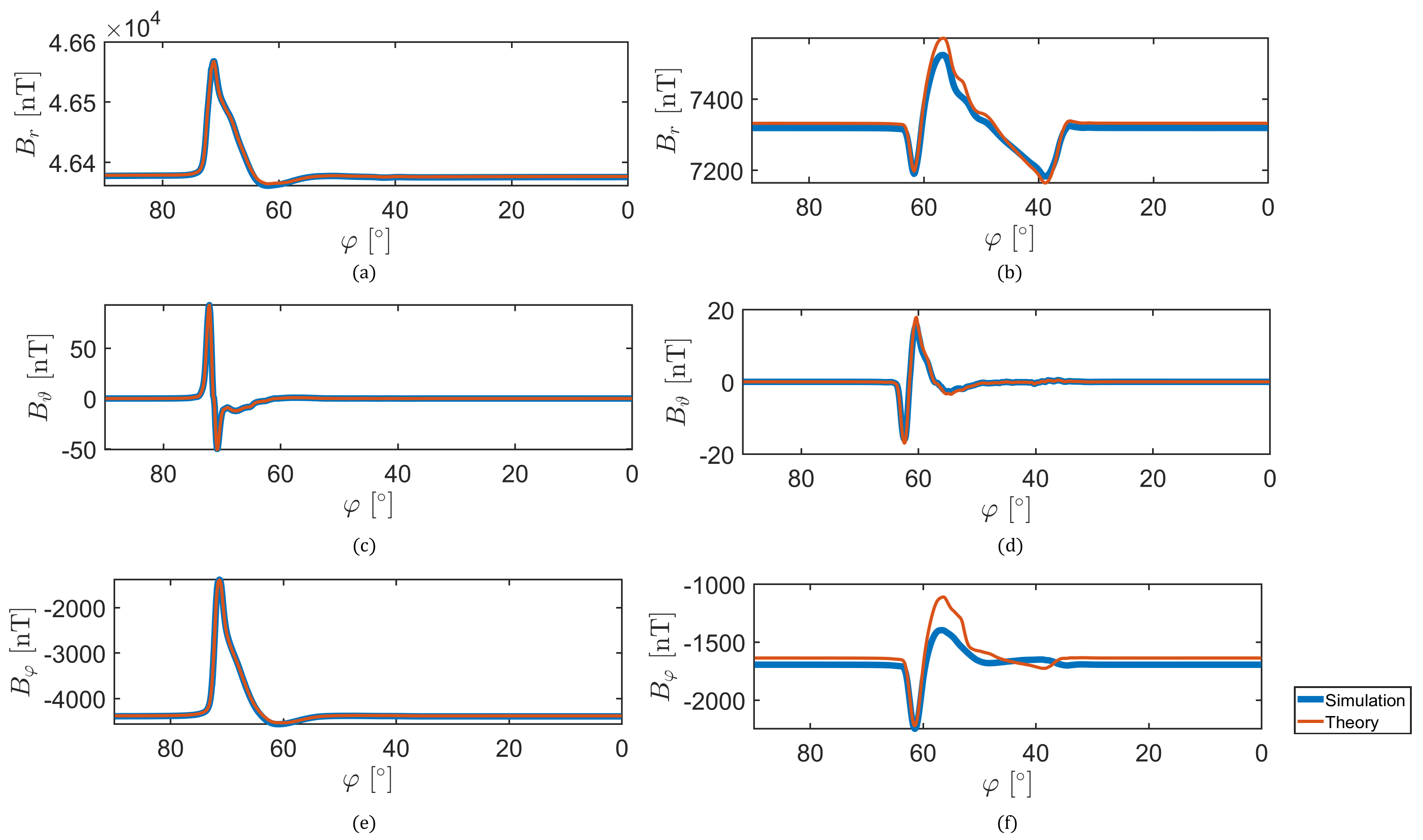}
\caption{Semi-analytic analysis of the Alfvén wings of TRAPPIST-1 b with line cuts along the $\varphi$-direction that display the magnetic field components $B_r$, $B_\vartheta$ and $B_\varphi$ (from top to bottom). The blue line represents the field component extracted from the simulation results and the red line shows the same components calculated with the Alfvén wing polarization relations (equations \ref{equation:AWnonlinBperp} and \ref{equation:AWnonlinBparallel}) and the simulated plasma velocity. The panels (a), (c) and (e), show a cut through the inward going Alfvén wing at $r=16\,R_*$. The panels (b), (d) and (f) show a cut through the outward going wave structures at $r=40\,R_*$.}
\label{fig:AlfvenCuts}
\end{figure*}

The Alfvén wings of a planet are arguably the most important wave structures for SPI, because they transport energy towards the star. In section \ref{section:WavePattern} we saw that Alfvén waves are the only waves that move inward to the star and outward. Therefore, we analyse both positions. To show that the leftmost wave structure that is indicated by the characteristics is indeed an Alfvén wing, we compare semi-analytic expectations with actually simulated magnetic fields.  Figure \ref{fig:AlfvenCuts} shows line cuts of the magnetic field components. Blue curves show the simulated data and red curves show the theoretically expected perturbations based on the polarization relations of the Alfvén wing model, i.e. equations \ref{equation:AWnonlinBperp} and \ref{equation:AWnonlinBparallel} and the simulated plasma velocity. The left panels (a), (c) and (e) show cuts along the $\varphi$-direction at $r = 16 \, R_*$, i.e. for the inward going Alfvén wing. The panels (b), (d) (f) on the right, show cuts along the $\varphi$-direction at $r = 40 \, R_*$, i.e. for the outward going Alfvén wing. The top row with panels (a) and (b) shows the $B_r$ component, the middle panels (c) and (d) show the $B_\vartheta$ component and the bottom row with panels (e) and (f) shows $B_\varphi$. Note that we reversed the $\varphi$-axis, to allow a more intuitive comparison with the plots of the equatorial plane.

We will first focus on the inward going Alfvén wing. The Alfvén wave produces perturbations in all three magnetic components. The strongest perturbation occurs in $B_\varphi$ with $2500$ nT. The main reason for that is that the planet moves nearly perpendicular to the plasma flow and the magnetic field into the $\varphi$-direction. The second strongest perturbation occurs in $B_r$ with $200$ nT. Although the radial magnetic field is the clearly dominating component in our stellar wind, the magnetic field has a slight curvature due to the stellar rotation. Therefore, the Alfvén wave, which only generates perturbations that are not parallel to the background magnetic field, also generates a perturbation in radial direction. The $B_\vartheta$ component exhibits the weakest perturbation with $70$ nT. However, since the background field has no $\vartheta$ component, this perturbation is easily visible from the background. The $\vartheta$-perturbation oscillates around zero. First the perturbation increases to 70 nT at about $72^\circ$ and then decreases to about $-40$ nT at $70^\circ$. We saw the same pattern in the $B_\vartheta$ component in Figure \ref{fig:Characteristics} (c). When we compare simulation with expectation, i.e. blue and red curve, we see excellent agreement between the two. This agreement indicates that, as expected, the inward going wave structure consists almost purely of Alfvén waves.

In contrast to the inward going Alfvén wing, we know from the analysis of the wave paths that the outward going Alfvén wing is one among many wave structures. Therefore, we can use the wave paths and the comparison to distinguish the Alfvén wave from the other waves. At $55^\circ$ the magnetic field exhibits a first perturbation that agrees well with the theoretical curve. The perturbation coincides with the Alfvén characteristic from section \ref{fig:Characteristics}. In $B_\varphi$ the perturbation makes about $-500$ nT. In $B_r$ it is $-150$ nT and in $B_\vartheta$ the wave produces a perturbation of about $10$ nT. The polarity of the wave is the opposite of the inward going wave. This anti-correlation agrees with theoretical expectations and is caused by the direction of propagation along the magnetic field line. The $B_\vartheta$ component is the only one that shows excellent agreement with the semi-analytic curve over all $\varphi$. The reason for that is that the other wave modes do not generate perturbations in $B_\vartheta$ and this component is only affected by the Alfvén wave. The other components, however, are subject to variations caused by other waves as well. We can see that in deviations between the blue and red curve. In $B_r$ there are only slight deviations, mainly at the strong peak at $~55^\circ$. The $B_\varphi$ component shows stronger deviations in the range of the other wave modes, i.e. between about $55^\circ$ and $40^\circ$.

\subsection{Slow Mode and Slow Shock \label{subsection:SloMoSloSho}}

\begin{figure}[t]
\centering
\includegraphics[width=0.5\textwidth]{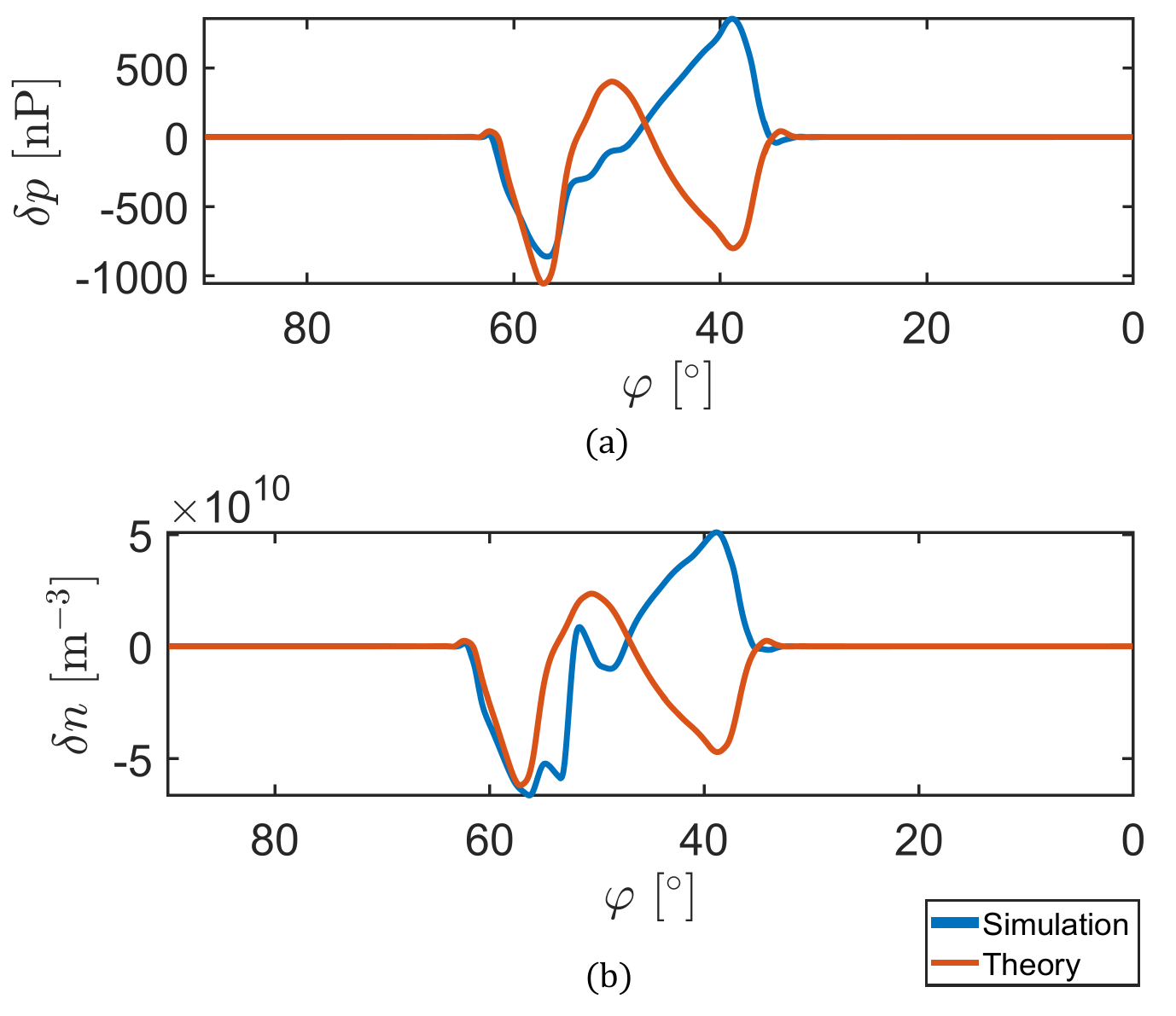}
\caption{Semi-analytic analysis of the Slow Mode of TRAPPIST-1 b with line cuts along the azimuthal direction at $r= 40 \, R_*$. Presented are the pressure perturbation $\delta p$ in panel (a) and the density perturbation $\delta \rho$ in panel (b). Blue lines represent simulated values and red lines are semi-analytic expectations based on linear wave theory.}
\label{fig:SlowStructures}
\end{figure}

\begin{figure*}[t]
\centering
\includegraphics[width=17cm]{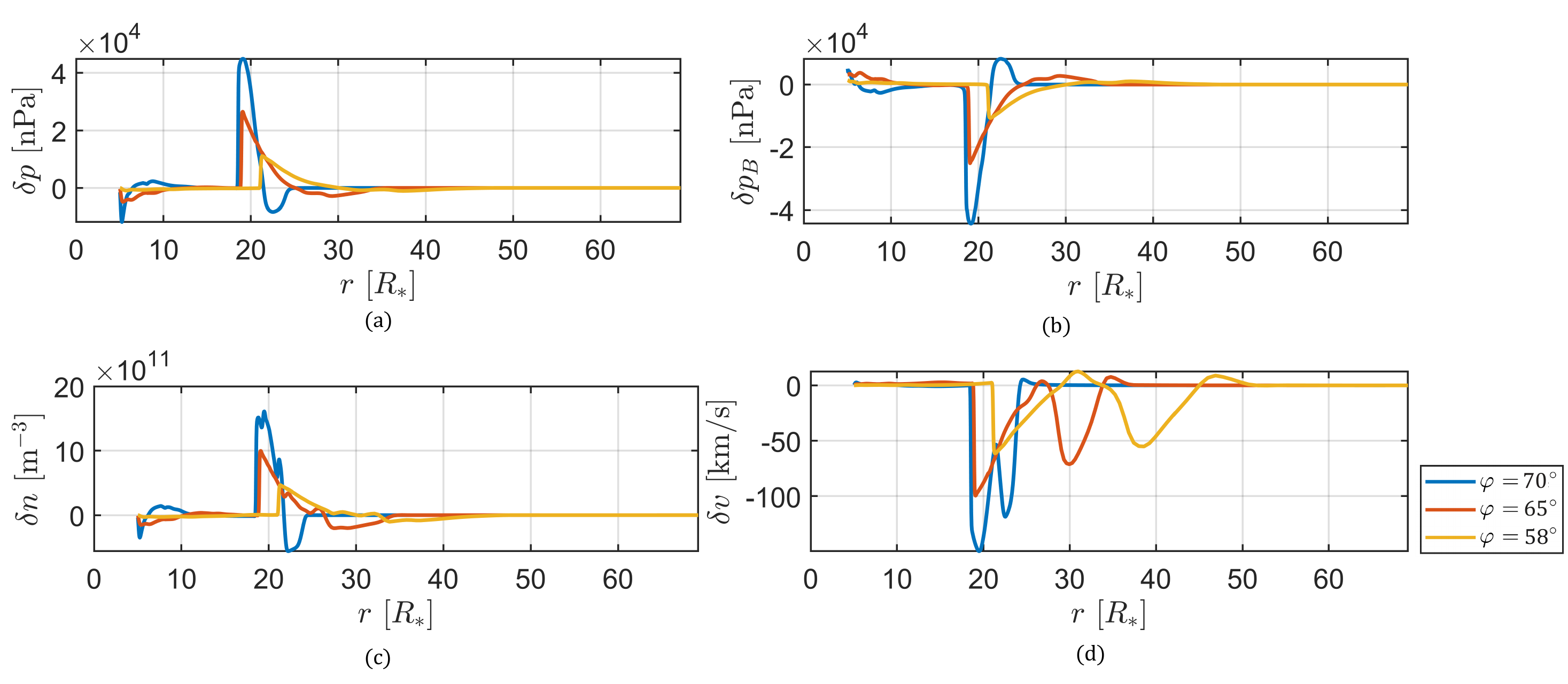}
\caption{Analysis of the Slow Shock of TRAPPIST-1 b with line cuts along the radial direction at three different azimuth angles through the outer wave structures of the planet. Presented are the pressure perturbation $\delta p$, magnetic pressure perturbation $\delta p_\mathrm{B}$, density perturbation $\delta \rho$ and the perturbation of the total velocity $\delta v$. \label{fig:RadialCuts}}

\end{figure*}

Due to the super-sonic interaction with the planet, one would expect a shock related to the slow mode. However, as we saw in \ref{section:WavePattern}, we obtain an outward going slow mode and a Shock that results from the inward going slow mode. In the following, we will conduct further analyses, similar to the Alfvén wings from section \ref{subsection:AlfvenWings}. However, for the slow mode, we rely on the linear wave theory from section \ref{subsection: MHDwaves} and compare theoretical expectations with actual perturbations in Figure \ref{fig:SlowStructures}. For the assumed shock, we have no theoretical expectation, so we look at cuts along the radial direction in Figure \ref{fig:RadialCuts} to identify the type of shock.

For the slow mode, we calculate the disturbances in pressure $\delta p$ and density $\delta \rho$ that arise from the velocity disturbance parallel to the background magnetic field with equations \ref{equation:SMp} and \ref{equation:SMrho} respectively. We compare them to the corresponding simulated disturbances. Figure \ref{fig:SlowStructures} shows the results for a distance of $40 \, R_*$ along the $\varphi$-direction. Blue shows the simulated values and red shows the perturbation of the slow mode based on the simulated plasma velocity to obtain an estimate for $\delta n$ and $\delta p$.

In section \ref{subsection:AlfvenWings}, we saw that the Alfvén wing lies at approximately $62^\circ$. According to our findings from section \ref{section:WavePattern}, we can expect the slow mode to be the next wave in clockwise direction, implying a lower value for $\varphi$ than for the Alfvén wing. The strongest negative peaks in $\delta p$ and $\delta \rho$ occur at $\varphi=57^\circ$. The location of the interaction peak is the same in the simulated values, as well as the semi-analytic values. However, there are deviations in strength. The actual perturbation of the pressure goes to $-800$ nPa and the semi-analytic expectation goes down to $-1000$ nPa. Similarly in $\delta n$. There, the simulation peaks at $-6\cdot 10^{10} \, \mathrm{m}^{-3}$ and the semi-analytic result lies at $-5.5\cdot 10^{10} \, \mathrm{m}^{-3}$. The deviations result from the linear wave theory, which assumes small disturbances. In our case here, we have strong interactions with an interaction strength (equation \ref{equation:InteractionStrength}) of about 0.9. That results in correspondingly large wave amplitudes of about 60\% of the local background pressure and density.

The wave-mode test with analytic expressions of the slow mode cannot explain the disturbances in pressure and density that appear at smaller $\varphi$. However, we already saw that these structures are connected to totally different waves. The last structure at $\varphi=39^\circ$ shows some anti-parallel correlation between both curves. This correlation may arise from the suspected slow shock. The narrow density peak at $\varphi=50^\circ$ will be subject of section \ref{subsection:MHDWake}.

When we look at figures \ref{fig:RadialCuts} (a) and (b), we see cuts of the thermal pressure disturbance $\delta p$ and the magnetic pressure disturbance $\delta p_\mathrm{B}$ along the radial direction at azimuthal positions $\varphi$ of $70^\circ$, $65^\circ$ and $58^\circ$. At $r=18\,R_*$ there are strong perturbations, which we will discuss later in this section. The perturbations at about $22\,R_*$, however, are connected to the outward going slow mode (also see Figure \ref{fig:Characteristics}). If we compare thermal pressure $p$ and magnetic pressure $p_B$, we see that the profiles of both pressure types are exactly mirrored counterparts of each other, which balances the total pressure in the plasma \citep{Linkerea88,Linkerea91}.

The wave structure behind the Alfvén wing is the outward going slow mode. The wave structure behind the wake is suspected to be a slow shock. Shocks are non-linear structures typically connected to plasma waves. The best known cases of shocks are the planetary bow shocks that form upstream of the planets in our Solar System. There, the planets experience super-fast plasma conditions, where no MHD wave can travel upstream towards the sun. In our case we have an intermediate situation where the plasma is sub-Alfvénic and super-sonic. Among the different types of shocks, the so-called slow shock \citep{BazerEricson59} is the best candidate to explain the occurring structure.

The different types of shocks all have their own distinct influence on the plasma. In principle, shocks are special kinds of discontinuities, where the plasma parameters change abruptly in form of a jump. The jump occurs in the direction of the plasma flow. In our case here, this is mainly the radial direction. Slow shocks form jumps in pressure, density, magnetic pressure and velocity \citep{BaumjohannTreumann96}. Pressure and density increase across the shock surface, while magnetic pressure and velocity will decrease. A Fast Shock, for example, causes a reduction in the magnetic pressure.

To identify the shock, we focus on the strong perturbations that appear in the radial profiles shown in Figure \ref{fig:RadialCuts}. Jumps in the direction of the flow describe the shock, which is nearly the radial direction in our case here. 

We can neglect the disturbances at small $r$, i.e. close to the star. Those result from the interaction between the inward going Alfvén wing with the inner boundary zone. We will focus on disturbances that appear at $r>18 \, R_*$, i.e. around the planetary orbit and outside. The shock causes jumps in pressure and density. Within about $0.3 R_*$, i.e. 8 grid cells, the pressure increases by $4.1 \cdot 10^{4}$ nPa (Figure \ref{fig:RadialCuts} (a)) and the number density increases by $15\cdot10^{11} \, \mathrm{m}^{-3}$ (Figure \ref{fig:RadialCuts} (c)) in the blue curve, i.e. at $\varphi=70^\circ$. When we compare pressure (subfigure (a)) and magnetic pressure (subfigure (b)), we see that both curves mirror each other. Therefore, the shock shows the properties of a slow shock. The other curves (red and yellow) represent former azimuthal positions of the planet and therefore show structures that were generated at earlier times $t$. There, we see that the amplitudes of the Slow Structures decrease as they propagate through the stellar wind. 

The jump in the blue and red curve already appears at $r=18\,R_*$, i.e. $2.5\,R_*$ within the orbit of planet b. For a magnetic field geometry that is almost parallel to the plasma flow, the slow mode cannot propagate against the super-sonic stellar wind towards the star. Velocity reductions connected to the Alfvén wing and possible small scale wave mode conversions cause locally reduced sonic Mach numbers. As long as the slow mode propagates with the guiding magnetic field and along the Alfvén wing, it can propagate towards the star. At some point this mechanism does not allow a further propagation because the stellar wind becomes super-sonic again. Then the slow mode forms a shock that propagates away from the star.

\subsection{Wake \label{subsection:MHDWake}} 
A planetary surface absorbs plasma, which causes a wake behind the planet. In the literature, planetary wakes are sometimes referred to as Entropy Mode \citep{JeffreyTaniuti64,Neubauer98,Saur21}. We saw a weak density disturbance in Figure \ref{fig:Characteristics} (b) that does not propagate by itself but is carried away by the stellar wind. This structure also appears in the radial profiles presented in Figure \ref{fig:RadialCuts} (c) that we already mentioned in section \ref{subsection:SloMoSloSho}. 

Naturally, one expects a decreasing density and pressure in the wake of a planetary object because the object absorbs plasma and shield the downstream region.  Comparing the wake path in Figure \ref{fig:Characteristics} (b) with the corresponding perturbations in Figure \ref{fig:SlowStructures} (b) and Figure \ref{fig:RadialCuts} (c) we see an increase in the density. In our simulation, there is no absorption of plasma, because the planet only reduces the plasma velocity through collisions. The deceleration causes a pile-up of plasma upstream of the planet. The planetary motion perpendicular to the stellar wind then directs the plasma around the planet and compressibility effects cause the wake that we see in the results. Due to the slow shock structure, the wake only plays a minor role in the wave pattern. Hence, only a small amount of density accumulates in front of the planet and eventually forms the wake structure.

\subsection{Impact on outer planets \label{subsection:ImpactOuterPlanets}} 

\begin{figure*}[t]
\centering
\includegraphics[width=17cm]{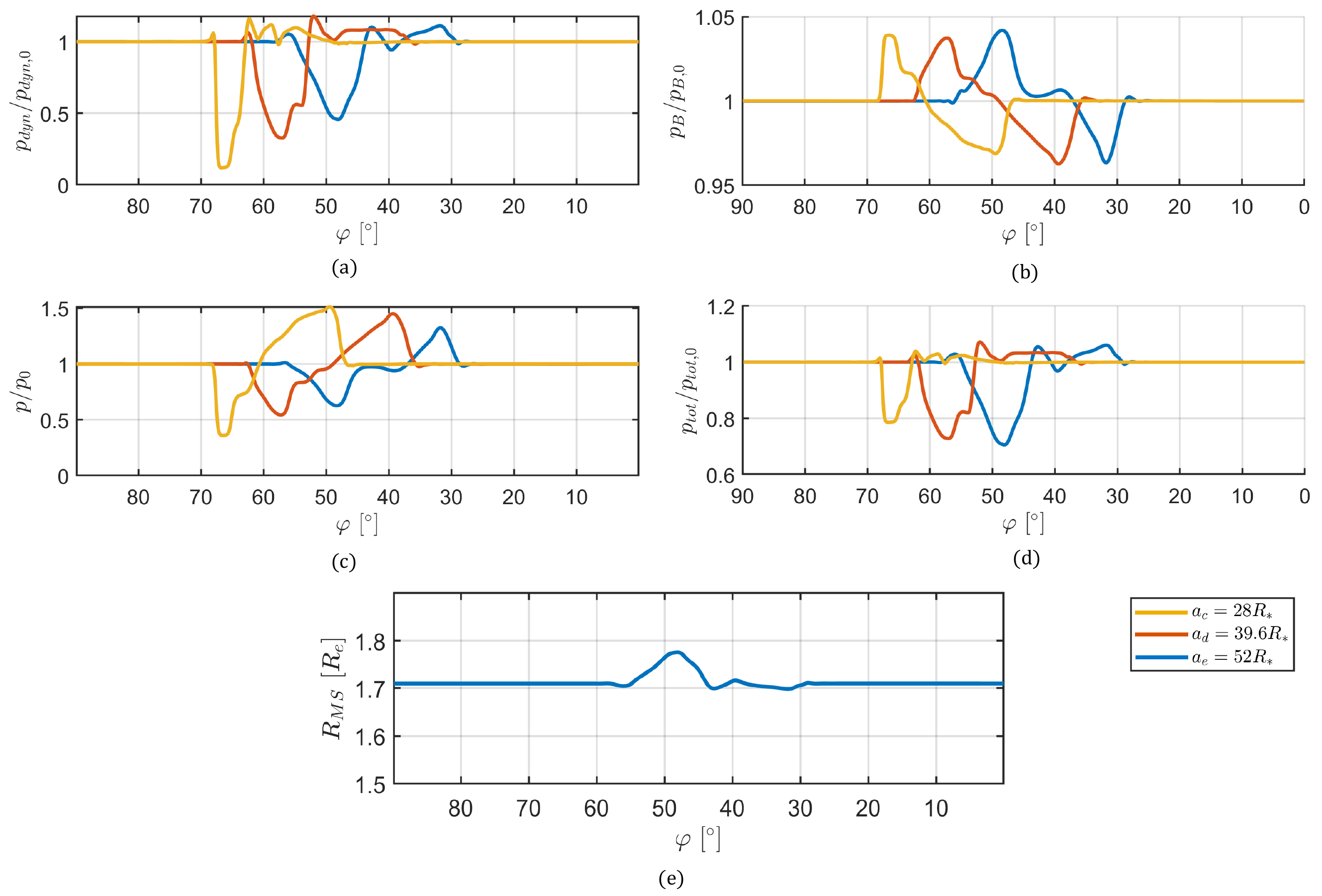}
\caption{Analysis of the impact of TRAPPIST-1 b's outward going wave structures on its neighbour planets. The panels (a) to (d) of this figure show the normalized dynamic, magnetic, thermal and total pressure respectively. Each line is cut along the azimuthal direction at the respective distances of the planets c (blue), d (red), and e (yellow). Panel (e) shows the hypothetical magnetospheric stand-off distance $R_\mathrm{MS}$ of planet e and how it is affected by planet b's wave structures.}
\label{fig:InfluenceOuterPlanets}
\end{figure*}

We saw that the wave structures originating from a planet can alter the space environment of planets that reside further outside (and inside) if the stellar magnetic field lines have a quasi-open geometry. To quantify the variation caused by SPI, we calculated the dynamic pressure $p_\mathrm{dyn} = \rho v^2$, the magnetic pressure $p_\mathrm{B} = B^2/(2 \mu_0)$, the thermal pressure $p$ and the total pressure $p_\mathrm{tot}$ as the sum of these three other pressures. We are interested in the variation that each of planet b's neighbouring planets may experience. All planets perform nearly circular orbits \citep{Grimmea18} and three of them, in addition to planet b, reside at distances that are included in our simulation box. Planet c (yellow) orbits TRAPPIST-1 at $r=28 \, \mathrm{R}_*$, planet d (red) at $r=39.6 \, \mathrm{R}_*$ and planet e (blue) at $r=52 \, \mathrm{R}_*$. Planet f would still lie within our simulation domain, but with an orbital distance of $r=68 \, \mathrm{R}_*$ it is very close to the outer boundary of the simulation box, which could include boundary effects. To compare these planets, we extracted the pressures along their orbits, i.e. in $\varphi$ direction, and normalized each pressure to the respective ambient pressure $p_0$ to allow a comparison between all planets. Since the stellar wind is uniform along the $\varphi$-direction, we can extract the normalization pressure $p_0$ from a region along the orbit that has not experienced any perturbations from SPI. Figure \ref{fig:InfluenceOuterPlanets} (a) to (d) shows the results. Panel (a) shows the normalized dynamic pressure, panel (b) the normalized magnetic pressure, (c) the normalized thermal pressure and (d) the normalized total pressure. 

The dynamic pressure $p_\mathrm{dyn}$ shows the largest overall variations compared to the other pressures. At planet c, $p_\mathrm{dyn}$ varies by about $90\%$. Whereas, the relative amplitude declines with distance from the star. So at planet e, the variability is $70\%$ of the ambient pressure. The major initial decline in the dynamic pressure is caused by the outward going slow mode. The remaining wave structures cause perturbations of about $10\%$ in the dynamic pressure. The magnetic pressure $p_\mathrm{B}$ shows the weakest relative perturbations with variation of about $5\%$ at all investigated planets. The thermal pressure $p$ in contrast shows similarly strong variations as the dynamic pressure, with amplitudes between $-70\%$ and $+50\%$ at planet c. Similar to the dynamic pressure, the variations decrease with distance, i.e. about $30\%$ at planet e. The decline at larger $\varphi$ in the pressure comes from the slow mode, while the increase at lower $\varphi$ is due to the slow shock.

All three pressures combined yield the total pressure $p_\mathrm{tot}$. This pressure shows variations between about $10\%$ at planet c and $25\%$ at planet e. Overall, the amplitudes are smaller than, for example in the thermal and dynamic pressure. The reason for this discrepancy is that the magnetic pressure is much larger than the thermal pressure. The thermal pressure accounts for maximum $10\%$ of the total pressure, whereas the contribution of the magnetic pressure ranges from nearly $100\%$ close to the star to $42\%$ at the outer boundary. At the same time, the dynamic pressure increases with radius. That explains the increase in the relative variation with distance from the star. Furthermore, the variations in the total pressure resemble the shapes that we saw in the dynamic pressure. That is a result of the connection between thermal and magnetic pressure. We already saw that in section \ref{section:WavePattern}. Those two pressures balance each other. The dynamic pressure is not involved in this relation because it is actually exerted by the flow of the fluid.  Therefore it affects planetary magnetospheres.

A planet's magnetosphere is basically a magnetic cavity within the stellar wind. As such, the magnetospheric stand-off distance is defined as the equilibrium point between the magnetic pressure of the intrinsic planetary magnetic field and the surrounding pressure. In the solar wind, one typically assumes the dynamic pressure of the solar wind. However, in the case of massive and strongly magnetized winds from M-dwarfs, one has to include the ambient magnetic pressure and the thermal pressure as well resulting in the total pressure $p_\mathrm{tot}$ that acts on planetary magnetospheres. Assuming, that the planetary field is a dipole field and other pressure contributions are negligible, we can estimate the magnetopause stand-off distance as
\begin{equation}
R_\mathrm{MS} = R_\mathrm{p} \sqrt[6]{\frac{B_\mathrm{p}^2}{2 \mu_0 p_\mathrm{tot}}}.
\end{equation}
The relation gives the magnetopause distance  and involves the planetary magnetic field strength at the equator $B_\mathrm{p}$ and the planetary radius $R_\mathrm{p}$.

TRAPPIST-1 e is a possibly interesting example for the effect that environmental changes imposed from the SPI wave structures can have. The fourth planet of the TRAPPIST-1 system orbits the star at a distance of about $52 \, R_*$ with an orbital period of 6.1 days. At the same time, planet e has with $1.024 \, \rho_\mathrm{E}$ almost exactly the same density as Earth \citep{Grimmea18}, which makes it much denser than all the other planets of the system. That hints at the existence of a metallic core, that may generate an intrinsic magnetic field and, therefore, a magnetosphere.

We assume that planet e has an Earth-like magnetic field strength of $B_\mathrm{p} = 30\,000$ nT at its equator. Figure \ref{fig:InfluenceOuterPlanets} (e) shows the size variation of planet e's hypothetical magnetosphere along the planet's orbit due to planet b's wave structures. In the ambient stellar wind, we see that the obstacle formed by the magnetosphere has a radius of $R_\mathrm{MS} = 1.7 \, R_\mathrm{e}$, with the radius $R_\mathrm{e}$ of TRAPPIST-1 e. Within the wave structures of planet b, at around $50^\circ$, the size increases by $0.1 \, R_\mathrm{e}$. Therefore, the cavity formed by the magnetic field varies by 15\%, just from the influence of planet b's wave structures. 

Past studies have focused on the influence of the stellar winds on exoplanetary magnetospheres, e.g. for planets around M dwarfs\citep{Vidottoea13,Vidottoea14}, for Hot Jupiters in super-Alfvénic stellar winds \citep{Vidottoea15} and for the TRAPPIST-1 system \citep{Garraffoea17}. \cite{Vidottoea14} found that the sizes of planetary magnetospheres around M dwarfs vary by about $20\%$ along the planetary orbits. \cite{Garraffoea17} conducted a parameter study for the planets of the TRAPPIST-1 system that is similar to our study. The authors assumed equatorial planetary magnetic field strengths of $10\,000$ nT and $50\,000$ nT and stellar magnetic field strengths of $0.3$ T and $0.6$ T. The case that best resembles our model case is the one with a stellar magnetic field of $0.6$ T. For TRAPPIST-1 e \cite{Garraffoea17} estimate a magnetosphere size of about $1.3$ and $1.8 \, R_\mathrm{e}$, depending on the assumed magnetic field. These estimates are very similar to our estimate with $1.7 \, R_\mathrm{e}$. \cite{Garraffoea17} assumed a more complex magnetic field geometry and can therefore see variations due to the stellar current sheet. The current sheet causes variations of about $0.1 R_\mathrm{e}$ of the magnetopause distance. Therefore, our results show that SPI can have a similarly strong influence on planets' magnetospheres as large scale stellar wind structures.

\section{Wing-Wing Interaction \label{section:WingWing}}
Wing-wing interaction is one of the four mechanisms described by \cite{FischerSaur19} that causes a time-variability in SPI. Understanding how the process works may have implications for observations of SPI. We simulate a hypothetical scenario involving the planets TRAPPIST-1 b and c. According to \cite{FischerSaur19} those planets are the most likely candidates for hosting SPI in the TRAPPIST-1 system. To understand the interaction better, we will compare the interaction of both planets first. Then, we will analyse the temporal evolution of the Alfvén wings and the compressional wave modes.

\subsection{SPI from two Planets \label{subsection:TwoPlanetSPI}} 

\begin{figure*}[t]
\centering
\includegraphics[width=17cm]{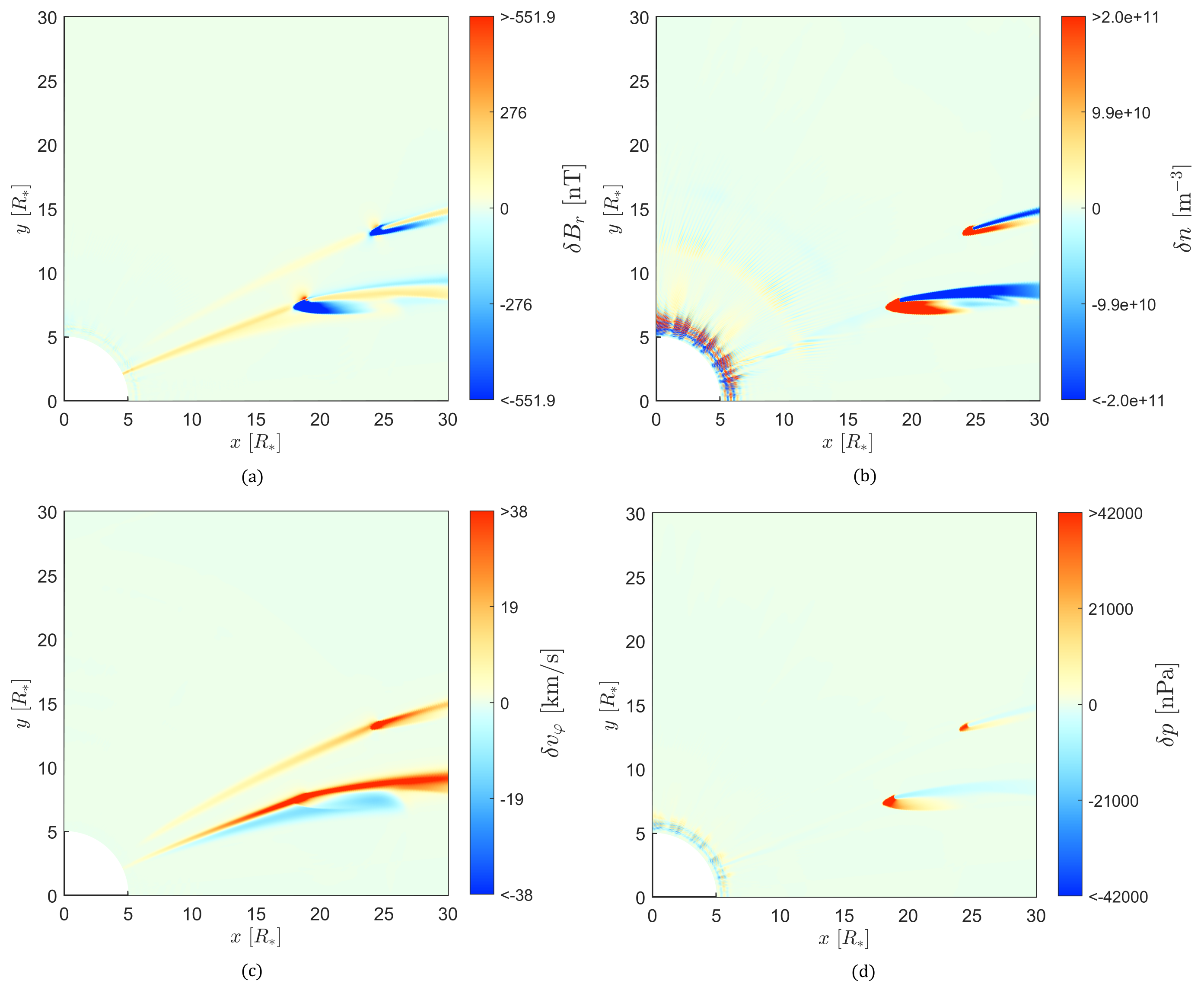}
\caption{Comparison of the interaction of planets b and c at $t=0.8$ hr. The panels shows the equatorial plane with $\delta B_\mathrm{r}$ in panel (a), $\delta n$ in (b), $\delta v_\varphi$ in (c) and $\delta p$ in panel (d).}
\label{fig:TwoPlanetSPI}
\end{figure*}

\cite{FischerSaur19} estimated that we can expect the power carried by the Alfvén wing of TRAPPIST-1 c to be about one order of magnitude smaller than the one generated by TRAPPIST-1 b. Accordingly, the power generated by TRAPPIST-1 d is already two orders of magnitude smaller than the one from TRAPPIST-1 b. Therefore, it becomes increasingly difficult to distinguish possible SPI-related signals of planets further away from the star from the strong signals coming from the close-in planets. Therefore, TRAPPIST-1 b and c are good examples for first studies of this kind, since they allow a better comparison.

For a comparison, we show the SPI created by both planets in Figure \ref{fig:TwoPlanetSPI}. We decided to discuss the results from the time $t=0.8$ hr, because the Alfvén wing of planet c arrived at the inner boundary. At this time the planets are still sufficiently far apart that they do not yet affect each other. Furthermore, both inward going Alfvén wings reached the inner boundary, which allows a proper comparison of the potentially interacting wave structures. 

The figure presents the perturbations in the radial magnetic field $\delta B_\mathrm{r}$, the number density $\delta n$, the azimuthal velocity $\delta v_\varphi$ and the pressure $\delta p$. Radial magnetic field and density have been chosen to compare this simulation to the example of a single planet from section \ref{section:WavePattern}. Azimuthal velocity and pressure are ideal variables to show temporal variability due to wing-wing interaction. We re-scaled the colorbar ranges of the presented disturbances to $\delta B_\mathrm{r} = \pm 551.9$ nT, $\delta n = \pm 2\cdot 10^{11} \, \mathrm{m}^{-3}$, $\delta v_\varphi = \pm 38 \, \mathrm{km \, s}^{-1}$ and $\delta p = 42\,000$ nPa, to enhance the visibility of the wave structures. The respective choices are determined by the strength of the interaction of the outer planet. The perturbations very close to the inner boundary that we see in the pressure and the density are boundary effects and have no physical origin.

We will first focus on the interaction of planet c. Figures \ref{fig:TwoPlanetSPI} (a) and (b) show that planet c produces qualitatively the same wave structures as planet b (see section \ref{section:WavePattern}). In the radial magnetic field perturbation $\delta B_\mathrm{r}$ (Figure \ref{fig:TwoPlanetSPI}(a)), we see the Alfvén wing that goes inward, although it is weaker than the one from planet b. In the azimuthal velocity perturbation $\delta v_\varphi$ (Figure \ref{fig:TwoPlanetSPI}(c)), we see that the Alfvén wing of planet c carries a perturbation of about $20 \, \mathrm{km \, s}^{-1}$ close to the planet. This value decreases towards the star. At planet b, the pattern is the same, but the Alfvén wing has a perturbation of $75 \, \mathrm{km \, s}^{-1}$ close to the planet.

Planet c also produces outward going waves, similar to planet b. At the chosen time, the Slow Structures and the wake of both planets only propagated about $10 \, R_*$ away from their planet. The respective travel times from the outer planet towards the outer boundary of the simulation domain ranges from $1.4$ hr for the Alfvén wing to $4.7$ hr for the slow shock. Therefore, the respective waves could not yet evolve to the same extent that we analysed for the inner planet in section \ref{section:WaveStructures}. However, we can already take conclusions concerning the shape of the wave structures. Figure \ref{fig:TwoPlanetSPI} (d) shows the pressure perturbation $\delta p$

Pressure and density nicely show the Slow Structures. The outward going slow mode generates a negative pressure disturbance and the slow shock generates a positive pressure disturbance. In accordance with our findings from section \ref{section:WaveStructures}, we see that the outward going slow mode is the leading wave structure here and the slow shock is the trailing structure relative to the planet's direction of motion. There are also some differences between both planets. First of all, we see that planet c generates wave structures that are much more confined than the ones generated by planet b. This change in the behaviour of the wave structures appears within a distance of $10 \, R_*$ between the inner planet and the outer planet. We attribute this change to the reduced wave speeds that do not allow the waves to expand too much. Furthermore, the stellar wind accelerates by about $20 \, \mathrm{km \, s}^{-1}$, which is a change of approximately 10\%. Both effects indicate an increasingly dominant radial component of the wave paths, which prevents a further broadening of the waves. In contrast to the inner planet, where we see very broad wave structures even close to the planet.

\subsection{Temporal Evolution of the Wave Structures} 
We will focus on the temporal evolution of the different wave structures, during the mutual interaction between both planets. We start with the Alfvén wings and proceed to the compressional wave modes.

\begin{figure*}[t]
\centering
\includegraphics[width=16cm]{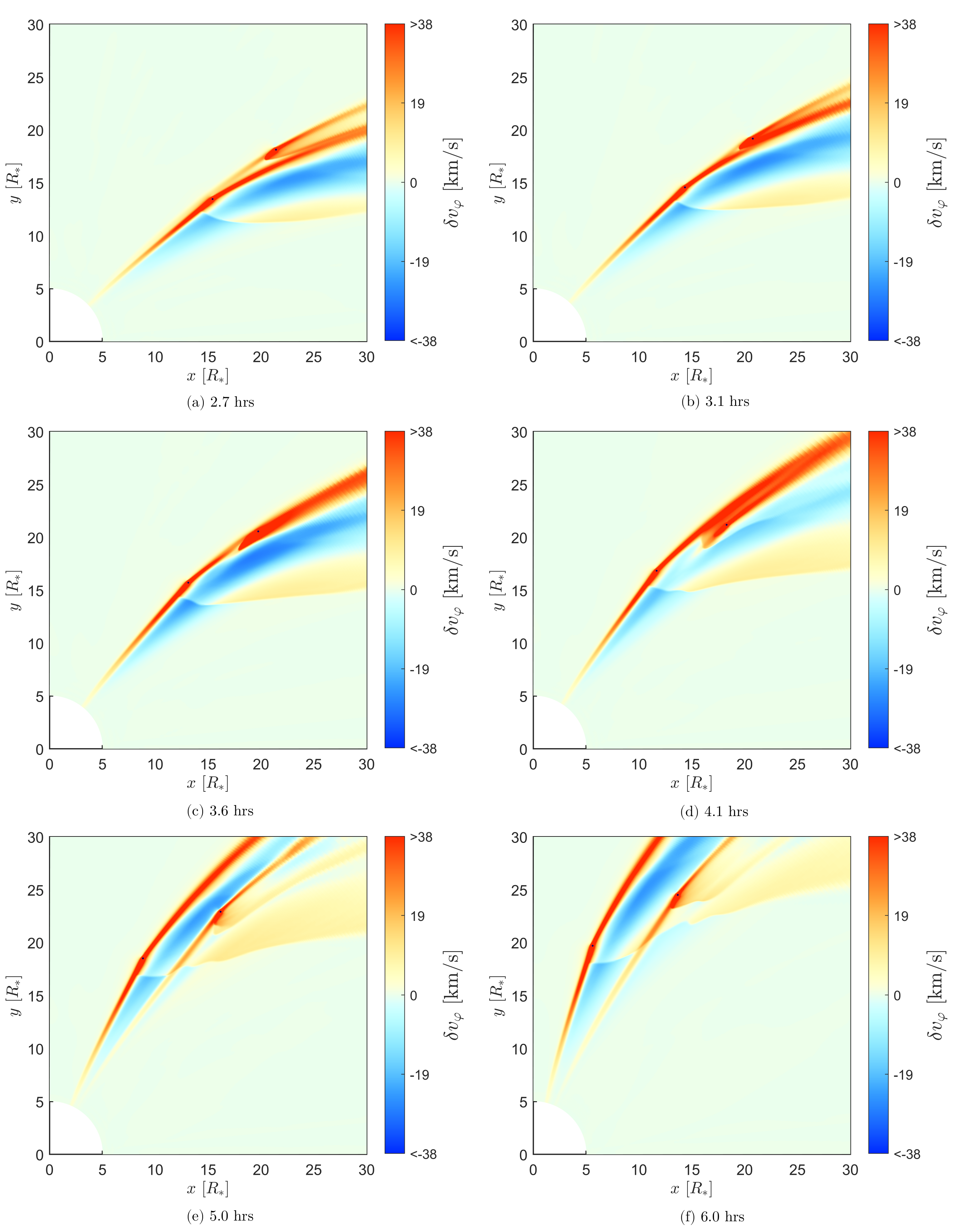}
\caption{Evolution of the Alfvén wings during the wing-wing interaction. The time range goes from $t=2.7$ hrs to $6.0$ hr. We show the evolution with the example of $\delta v_\varphi$. The colorbar is set to a range of $\pm 38 \ \mathrm{km \, s}^{-1}$, which is about half the maximum disturbance. The planets are indicated by blue dots that represent their actual size. See \textbf{Online movie} for the interaction and a comparison between $\delta v_\varphi$ and $\delta p$.}
\label{fig:TemporalEvolution1}
\end{figure*}

\subsubsection{Alfvén wings \label{sec:WWAlfvenWings}}

In this section, we will look at the temporal evolution of the wing-wing interaction between planet b and c. Figure \ref{fig:TemporalEvolution1} shows selected times during the wing-wing interaction for $\delta v_\varphi$. Panel (a) presents the situation at $t=2.7$ hrs, panel (b) at $3.1$ hrs, (c) at $3.6$ hrs, (d) at $4.1$ hrs, (e) at $5.0$ hrs and panel (f) at $t=6.0$ hrs. These times show the evolution of the Alfvén wing from a stage where planet b is just about to overtake planet c to a situation, when planet c is about to exit the wave structures of planet b. For this analysis, the velocity components are more practical, because their reflections at the inner boundary change the sign and allow to distinguish the actual wave from reflections. The left panel in \textbf{Online movie} shows the evolution of $\delta v_\varphi$.

In subfigure (a) at $t=2.7$ hrs, planet b is about to overtake planet c and at $t=3.1$ hrs, planet c enters the outward going Alfvén wing of planet b. However, the inward going Alfvén wings already interact with each other. At this stage, we can not see any apparent changes in the interaction.

At $t=3.6$ hrs (subfigure \ref{fig:TemporalEvolution1} (c)), planet c resides within planet b's outward going Alfvén wing. Hence, counterpropagating Alfvén waves interact with each other. The result is a considerable weakening of the inward going Alfvén wing of planet c. We see that at $t=4.1$ hrs in subfigure (d), the Alfvén wing is very weak and only reaches a few stellar radii towards the star. The Alfvén wing of planet c vanished through the interaction with planet b's strong outward going Alfvén wing and at $t=3.6$ hrs it emerges again.

At the later times, i.e. $t=5.0$ hrs and $6.0$ hrs, the planet c resides within the compressional wave modes of planet b. At these times we see a newly formed Alfvén wing from planet c. Since the Alfvénic travel time from planet c towards the inner boundary is about 0.8 hrs, planet c could now fully re-establish the coupling to the star. We know that the Slow Structures directly affect the density of the plasma and indirectly the magnetic field strength as well. Both quantities determine the Alfvén wave speed and therefore affect the path of the outer planet's Alfvén wing. Accordingly, the new Alfvén wing assumes a larger angle to the unperturbed magnetic field than it would have in the steady state stellar wind.

\subsubsection{Compressional waves}

\begin{figure*}[t]
\centering
\includegraphics[width=0.9\textwidth]{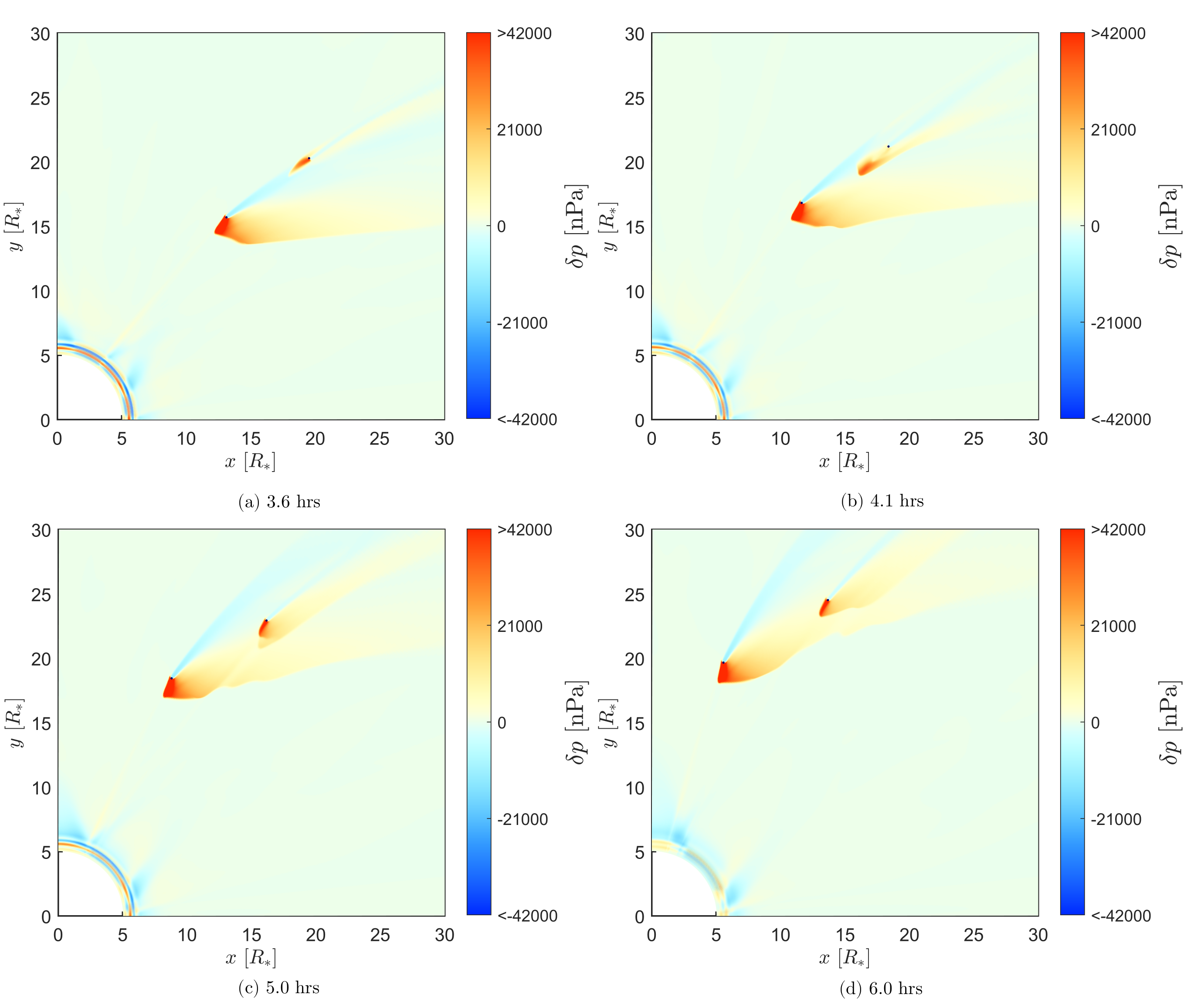}
\caption{Evolution of the Compressional wave modes during the wing-wing interaction. The time range goes from $t=3.6$ hr to $6.0$ hr. We show the evolution with the example of $\delta p$. The planets are indicated by blue dots that represent their actual size. See \textbf{Online movie} for the interaction and a comparison between $\delta v_\varphi$ and $\delta p$.}
\label{fig:TemporalEvolutionPrs}
\end{figure*}

To analyse the behaviour of the Slow Structures, we look at the temporal evolution of the pressure perturbation $\delta p$. In Figure \ref{fig:TemporalEvolutionPrs} we see the times from $t=3.6$ hrs to $6.0$ hrs because only these are relevant for the compressional wave modes. The planets are indicated by blue dots that represent their actual size. Compared to the situation at $t=0.8$ hrs, the Slow Structures evolved and extend much further. At $t=3.6$ hrs, we can now clearly see that the wave structures of planet c are much more confined than the ones originating from planet b (also see section \ref{subsection:TwoPlanetSPI}). The right panel in \textbf{Online movie} shows the evolution of $\delta p$.

After planet c leaves planet b's Alfvén wing, the Slow Structures of planet b directly affect planet c. There, the planet experiences a different plasma environment, which affects the generated waves. From section \ref{subsection:ImpactOuterPlanets} and Figure \ref{fig:InfluenceOuterPlanets} we know that planet c experiences pressure variations within the Slow Structures of planet b. The dynamic pressure varies by about 90\% and the thermal pressure by 60\% from the quiet state. Accordingly, the sound speed within planet b's slow mode is smaller than the ambient sound speed. Similarly, the plasma speed is lower in both Slow Structures (see section \ref{subsection:SloMoSloSho}). This affects the wave paths and allows broader wave structures. We see the result in Figure \ref{fig:TemporalEvolutionPrs} (c) and (d). At $t=3.6$ hrs the Slow Structures of planet c had a width of roughly one planetary diameter. At $t=4.1$ hrs the slow shock looks smeared out and at later times it reaches several planetary radii in width.

The interaction of planet c's Alfvén wing and the inner planet's slow shock produces wave-like curves appearing at the shock front (Figure \ref{fig:TemporalEvolutionPrs} (c) and (d)). It is possible that these perturbations of the shock front result from the velocity disturbances connected to the Alfvén wing because they alter the characteristics. Accordingly, the shock front follows the new path determined by the characteristics.

\subsection{Energetic Evolution}

\begin{figure*}[t]
\centering
\includegraphics[width=0.8\textwidth]{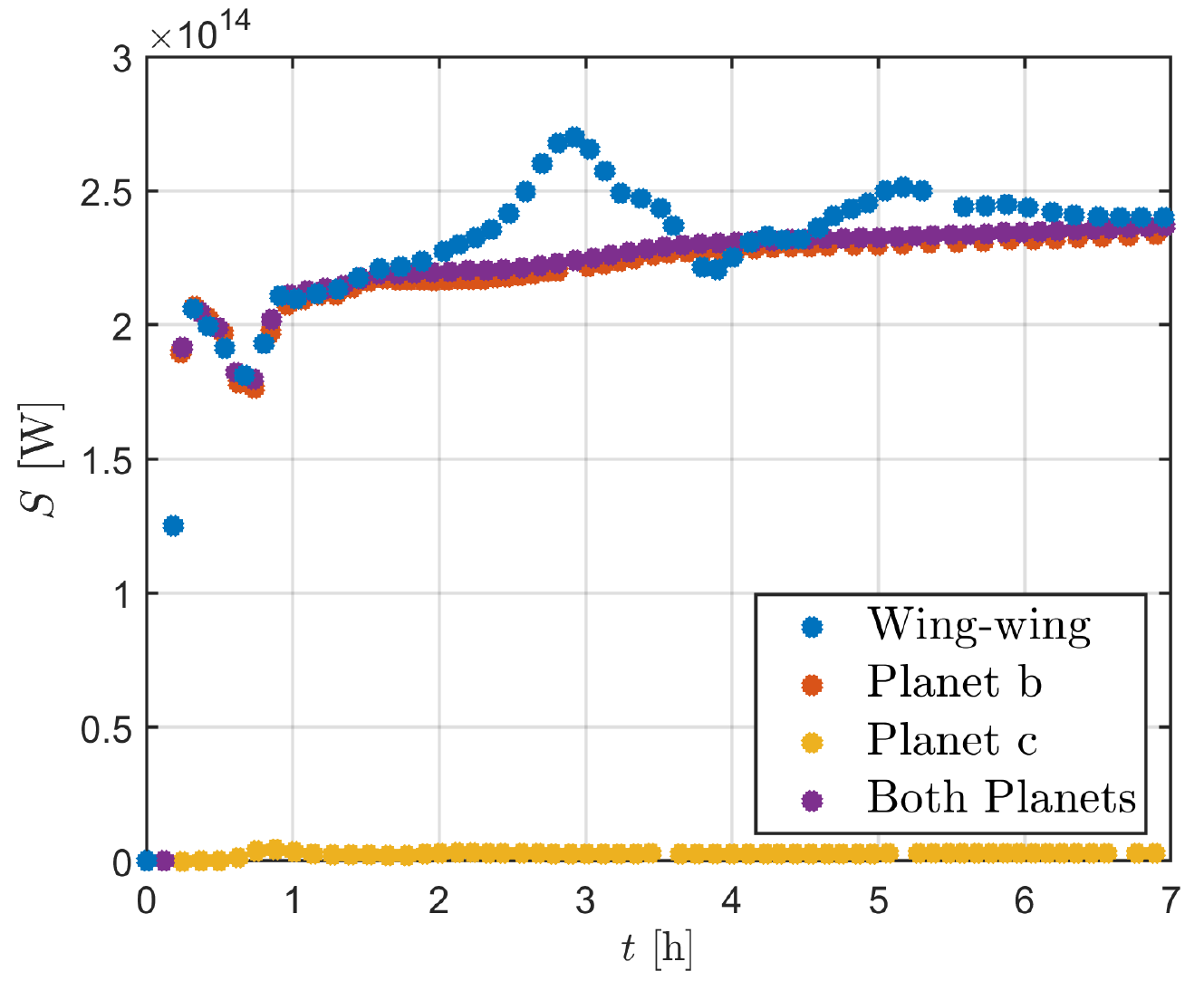}
\caption{The Poynting flux $S$ directed towards the star at $r=16 \, R_*$ generated during the wing-wing interaction (blue), compared to the Poynting flux of planet b (red), planet c (yellow) and the sum of both (purple). The time ranges from $t=0$ hrs to $7.0$ hrs.}
\label{fig:WWPoyntingFlux}
\end{figure*}

The Poynting flux determines the observability of SPI against the stellar background. We will now look at the kind of variability that wing-wing interaction will cause. To determine the Poynting flux that each Alfvén wing carries towards the star, we follow the method of \cite{Saurea13}, which is outlined in appendix \ref{section:Appendix}.

We calculated the Poynting flux $S$ as a function of time $t$ from $t=0$ hrs to $7$ hrs and displayed the result in Figure \ref{fig:WWPoyntingFlux}. The blue dots show the Poynting flux of our simulation with occurring wing-wing interaction. As a comparison, we carried out separate simulations for only planet b (red) and only planet c (yellow), and display their Poynting flux along with the sum from the individual runs (purple). Without non-linear interaction, i.e. wing wing interaction, the sum of these two Poynting fluxes should be similar to the joint simulation including both planets. All fluxes are determined at a radial distance of $r=16 \, R_*$ from the star. At that distance, we see both Poynting fluxes that go towards the star and have no effects from the compressional wave modes. When interpreting the results, we have to know that the Alfvén waves need about $0.5$ hrs to travel from planet c to the analysis distance at $r=16 \, R_*$ and that Alfvén waves originating at planet b need about $0.1$ hrs. These values were estimated for the unperturbed stellar wind. If planet c resides within the wave structures of planet b, there may be deviations in the travel time.

In the beginning, from $t=0$ hr to about 1 hour, the Alfvén wings of the planets form and go towards the star. Therefore, the Poynting flux ramps up until it reaches a quasi-stationary value. For planet b that value lies at about $S = 1.7 \cdot 10^{14}$ W and for planet c, at this distance, it is at about $S = 3.5 \cdot 10^{12}$ W. Both Alfvén wings lost small amounts of their power at this distance, mainly due to numerical dissipation. For planet c, the dissipation is stronger, because it already travelled a longer way. We further see that the Poynting flux of planet b increases slightly over time. We explain this effect with a weak non-stationary evolution of the stellar wind. 

Essentially, the time range from $0.9$ hrs to $1.7$ hrs shows a good overlap of the blue and the purple curve. The purple curve is just the sum of the separate Poynting fluxes of planet b and c. That shows that the simulation with both planets provides the same results as separate simulations. The situation changes at around $t=2$ hrs, when the blue curve clearly departs from the purple curve. We have designed the simulation such that both planets are in conjunction at $t=2.5$ hrs. Therefore, we expected the first signs of wing-wing interaction at around that time. Shortly after the conjunction at $t=3$ hrs, planet c enters planet b's outer Alfvén wing. That is the time, when the Poynting flux reaches its maximum of $S=2.7 \cdot 10^{14}$ W. The Poynting flux increases by about $20\%$ compared to a pure superposition of both planetary Poynting fluxes (purple curve). Afterwards, the Poynting flux decreases again, even below the sum of both separate Poynting fluxes. This local minimum at $t=3.7$ hrs takes about 20 minutes, then the Poynting flux increases slightly towards a local maximum of $S=2.5 \cdot 10^{14}$ W at $t=5.3$ hrs, which declines slowly until about $t=7$ hrs, when the wing-wing interaction is over and the planets only interact with the stellar wind again.

\begin{figure*}[t]
\centering
\includegraphics[width=1.0\textwidth]{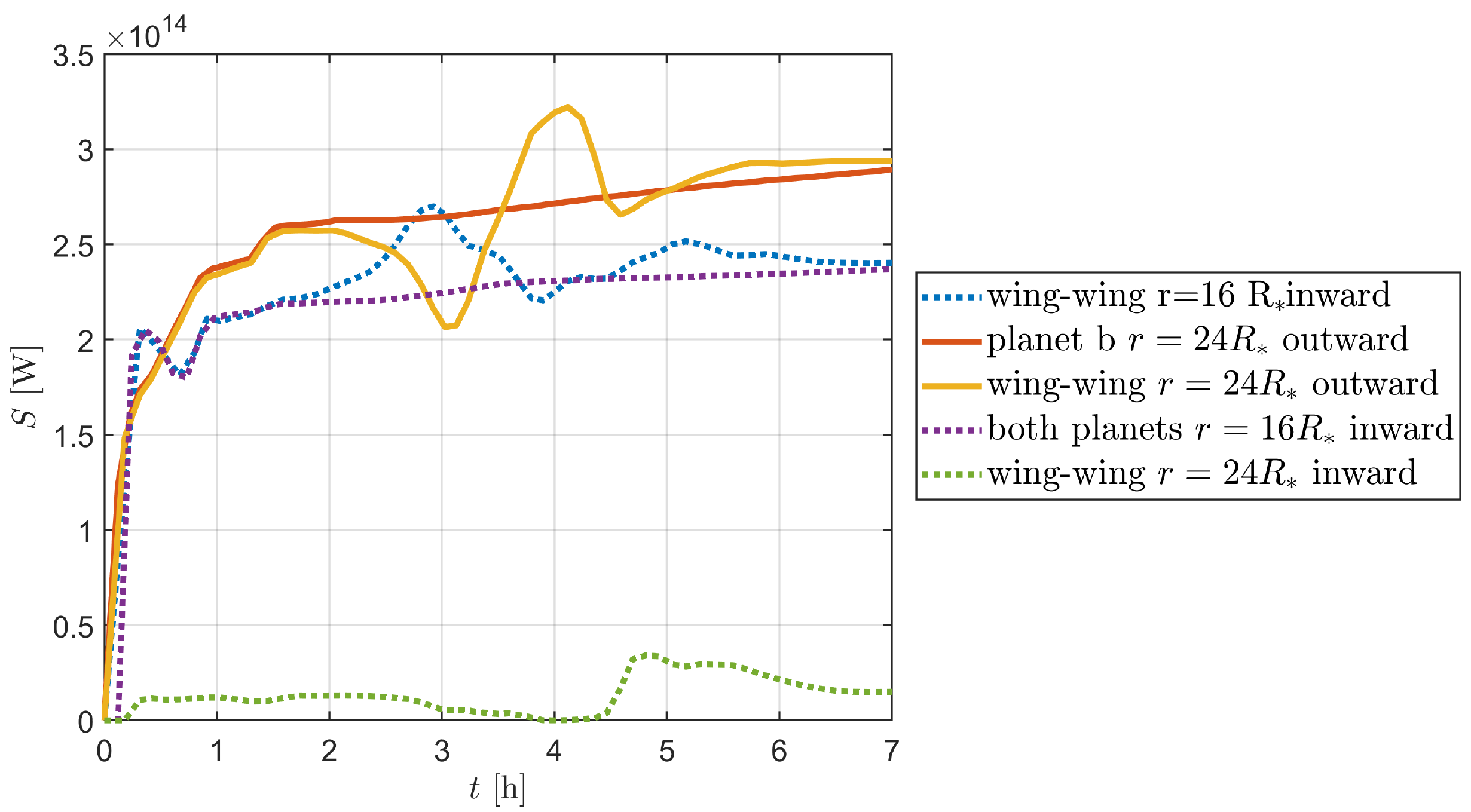}
\caption{The inward going Poynting flux $S$ generated during the wing-wing interaction (blue) together with the sum of the inward going Poynting fluxes from both planets (purple) evaluated at $r=16\,R_*$, compared to the outward going Poynting flux of planet b alone (red) and during wing-wing interaction (yellow) evaluated at $r=24\,R_*$. The time ranges from $t=0$ hrs to $7.0$ hrs.}
\label{fig:WWPoyntingFlux2}
\end{figure*}

To understand the origin of the variations in the Poynting flux, we look at the region between the two planets. There, the counterpropagating Alfvén waves from planet b and c interact with each other. Figure \ref{fig:WWPoyntingFlux2} shows the Poynting flux $S$ as a function of time $t$, just as Figure \ref{fig:WWPoyntingFlux}. Dashed lines show Poynting flux that goes towards the star and solid lines show Poynting flux that goes away from the star. The blue line shows the Poynting flux of the wing-wing interaction at a distance of $16 \, R_*$ and the purple line the summed up Poynting fluxes of the two planets. Those two lines are the same as in Figure \ref{fig:WWPoyntingFlux} and are included only for comparison. The yellow line shows the outward going Poynting flux of the wing-wing interaction at $24 \, R_*$, i.e. exactly between the two planets. The red curve shows the outward going Poynting flux of planet b alone for comparison. The green curve shows the inward going Poynting flux of the wing-wing interaction at $24 \, R_*$.

We see that the total outward going Poynting flux of the interacting Alfvén wings (yellow curve in Fig. \ref{fig:WWPoyntingFlux2}) reduces by up to $35\%$ at $t=3.1$ hrs compared to the non-interacting Alfvén wings (red curve). The peak in the total inward going Poynting flux (dashed blue curve) is the counterpart of the drop in the yellow curve and indicates the non-linear wave-wave interaction between the outward going Alfvén wing of planet b and the inward going Alfvén wing of planet c. The resulting daughter Alfvén waves propagate towards the star and receive their energy from the non-linear wave-wave interactions. The increase in the inward going flux and the decrease in the outward going flux at $t=3.1$ hrs occur when the Alfvén wings merge and interact with each other (see Figure \ref{fig:TemporalEvolution1} and \textbf{online movie}). Additionally, the blue curve features a special shape, with a small plateau-like phase at $t=3.5$ hrs. The yellow curve does not have such a plateau. We attribute this feature to planet b that blocks some of the waves and the Poynting flux that go towards the star. Accordingly, with an increase of $20\%$, the peak in the blue curve is weaker than the drop in the yellow curve with about $35\%$. The drop in the yellow curve and the peak in the blue curve, are shifted by about 10 minutes. We see the reason for this in the supplied \textbf{online video}. The merging zone between the Alfvénic wave structures of both planets starts close to the star and gets shifted outward. That is a result of the bent-back of the Alfvén wings within the stellar magnetic field. Thus outward going waves close to planet b interact much earlier with the inward going waves of planet c than the waves at $r=24 \, R_*$. The resulting daughter waves propagate towards the star and cause the peak at $r=16 \, R_*$  before wave-wave interactions occur at $r=24 \, R_*$.

The second peak in the blue curve at $t=5.3$ hrs is a result of the Slow Structures of planet b. The slow mode and the slow shock change the plasma properties around planet c (see section \ref{subsection:ImpactOuterPlanets}, Figure \ref{fig:TemporalEvolutionPrs} and \textbf{online movie}). At about $t=4.0$ hrs planet c leaves the outer Alfvén wing and enters the compressional wave structures. The first waves arrive at $r=16 \, R_*$ about half an hour later. Accordingly, the Poynting flux $S$ starts to increase at $t=4.5$ hrs and reaches its local maximum at $t=5$ hr. Looking at the slow mode structure, we see that the leading half of the wave structure carries stronger perturbations than the trailing side (Figure \ref{fig:TemporalEvolutionPrs}). Therefore, the Poynting flux decreases again, although the planet is still within the slow mode. At $t=5$ hrs, the planet enters the wake and the slow shock. This transition is visible in a small irregularity in the Poynting flux at $t=5.5$ hrs (Figure \ref{fig:WWPoyntingFlux}). However, the Poynting flux decreases further and eventually reaches the normal state where both planets interact with the stellar wind but not with each other. Figure \ref{fig:WWPoyntingFlux2} shows the correspondingly increased inward going Poynting flux at $24 \, R_*$ (green curve). Since planet c is the only planet that produces an inward going Poynting flux at this distance, we can attribute the green curve solely towards planet c and its interaction with the slow mode and the slow shock of planet b.

In total, we can say that the changes in the Poynting flux during the wing-wing interaction are a result of the non-linear interaction of Alfvén wings. Said interaction causes an increase of the wave perturbations and results in a larger Poynting flux between $t=2$ hrs and $3.7$ hrs. The increase of the Poynting flux between $t=4.5$ hrs and $7$ hrs is a result of planet c residing within planet b's Slow Structures.

\section{Conclusions \label{section:Conclusions}}
In this paper, we presented novel results of a fully time-dependent MHD model to simulate SPI in a multi-planet system. In the model, the planets are reduced to their atmosphere, which generates collisions between plasma ions and atmospheric neutral particles. The planets orbit their host star and are embedded in a Parker-like, thermally driven stellar wind with a quasi-open magnetic field geometry. Our results show that electromagnetic star-planet interaction is a process that affects planets as well as the host star.

In the first part of our analysis, we investigated and identified all wave structures that are related to SPI. This basic analysis is the first of its kind in the field of exoplanet research. Although these wave structures can not be observed directly, we show that they affect the space environment of other planets in a stellar system. The estimated perturbations of the thermal, magnetic and dynamic pressure in the stellar wind environment of the planets account for 10\% to 40\% variation relative to the quiet state. On the example of a hypothetical magnetosphere at TRAPPIST-1 e and in comparison to results from \cite{Garraffoea17}, we see that the effects of SPI compare to large scale variations in stellar winds caused by current sheets as well as slow and fast wind zones. Considerable changes in the small sizes of possibly existing magnetospheres would affect e.g. the auroral radio emission from the planet. 

Similarly to magnetospheres, SPI wave structures would affect and shape the atmospheres of unmagnetised planets. Especially the upper layers of the atmospheres and the gas coronae are heavily affected by stellar winds. Several studies investigated the role of stellar winds in the erosion of planetary atmospheres, e.g. \cite{RodriguezMoya19,VidottoCleary20} and \cite{Harbachea21}. Special interest lies in the absorption of Lyman-$\alpha$ in the gas corona of planets, so-called Lyman-$\alpha$ transits. This process is important to characterise atmospheric evolution through atmospheric escape \citep{Owenea21}. Tides, radiative pressure and the aforementioned stellar winds affect Lyman-$\alpha$ transits. Through the effect of SPI on stellar winds, one could expect an influence on Lyman-$\alpha$ transits. Despite, it is currently unknown how much the stellar winds of M-dwarfs vary on time scales of a few days. Therefore, it is not clear how the variability caused by the wave structures scales against the stellar wind variability . SPI may also affect the gas corona of the planet that generates it. Our results show considerable changes of the plasma environment around SPI-hosting planets. This could shape the atmospheric tails of planets and cause variations in Lyman-$\alpha$ transits.

Apart from the changes of the space environment, we saw that SPI wave structures can also interact with each other. Wing-wing interaction, i.e. the interaction between two Alfvén wings, is of interest because it can create observable features. We chose to investigate the planets Trappist-1 b and c, because their interaction produces comparable perturbations that only differ by a factor of 10. We found out that the total Poynting flux of both planets increases by up to $20\%$, when both inward going Alfvén wings merge. That is a result of the non-linear interaction of the wave perturbations. Our results show a distinctive time-variable pattern of the Poynting flux that would appear periodically with the synodic revolution period between both planets. An observation of such an energetic profile would provide indirect evidence for the existence of the compressional wave modes, especially slow mode and slow shock.

We can conclude that detailed knowledge of the phenomena that are connected to SPI are of special interest for exoplanet research. The electromagnetic connection between a star and its planet(s) is thus not just an issue that relates to stellar activity. In fact SPI appears to be a vital process for the whole stellar system.

\begin{acknowledgements}
C. F. and J. S. acknowledge funding by Verbundforschung
für Astronomie und Astrophysik through grant No. 50 OR 170. J.S. is supported by the European Research Council (ERC) under the European Union’s Horizon 2020 research and innovation programme (grant agreement No. 884711). We furthermore thank the Regional Computing Center of the University of Cologne (RRZK) for providing computing time on the DFG-funded (Funding number: INST 216/512/1FUGG) High Performance Computing (HPC) system CHEOPS as well as support.
\end{acknowledgements}

\bibliographystyle{aa}
\bibliography{My_Bib}

\appendix
\section{Estimating the planetary Poynting flux from the simulations \label{section:Appendix}}
The Poynting vector is the electromagnetic energy flux density that is carried by the Alfvén waves. Every plasma with a velocity field that is not parallel to the magnetic field carries such a Poynting vector. We refer to this flux as the background flux. However, \cite{Saurea13} provide the theory to obtain the energy flux that is solely carried by the Alfvén wing. Therefore, we have to apply a Galilei-transformation on the velocity field, to switch into the reference frame of the magnetic field. The transformation velocity $\textit{\textbf{u}}$ has to suffice the condition

\begin{equation}
\textit{\textbf{E}}_0 = (\textit{\textbf{v}}_0 - \textit{\textbf{u}} ) \times \textit{\textbf{B}}_0 = 0
\end{equation}

where the subscript 0 denotes the background values of the respective plasma variables. The radial stellar wind however has to remain unchanged because we will not switch into the system of the plasma. Hence, $\textit{\textbf{u}}$ has to fulfil the following conditions:

\begin{align}
\textit{\textbf{v}}_0 - \textit{\textbf{u}} & \parallel \textit{\textbf{B}}_0 \\
u_r & = 0.
\end{align}

To make the transformed velocity components in $\vartheta$ and $\varphi$ parallel to $\textit{\textbf{B}}_0$, we apply

\begin{equation}
\frac{v_\mathrm{i,0} - u_\mathrm{i}}{v_{r,0}} = \frac{B_\mathrm{i,0}}{B_{r,0}}
\end{equation}

with $i = \vartheta, \varphi$. The Poynting flux density carried by the Alfvén wing and in the inertial reference frame is then given by

\begin{equation}
s_\mathrm{AW} = \frac{ \left[ - \left( \textit{\textbf{v}} - \textit{\textbf{u}} \right) \times \textit{\textbf{B}} \right] \times \textit{\textbf{B}} }{\mu_0} \cdot \frac{\textit{\textbf{B}}_0}{|\textit{\textbf{B}}_0|}. \label{equation:AWPoyntingvector}
\end{equation}

The Poynting vector is an energy flux density, with units of W/m$^2$. To receive the Poynting flux $S$ transferred through the Alfven wing, i.e. the total power, we have to integrate $s_\mathrm{AW}$ over the cross-sectional area of the Alfven wing. We apply the Riemann integration formula over the $\vartheta$-$\varphi$-plane at a certain distance $r_\mathrm{eval}$ where we evaluate the power. The simulation grid provides the area elements $dA$ with 

\begin{equation}
dA = r_\mathrm{eval}^2 \sin(\vartheta) \, d \vartheta \, d \varphi.
\end{equation}

Accordingly, the simulated Poynting flux is given by

\begin{equation}
S = \sum_i \sum_j s_\mathrm{AW,ij} dA_\mathrm{ij} \label{equation:PoyntingPower}
\end{equation}

with the indices $i$ and $j$ representing the direction along $\vartheta$ and $\varphi$.

\end{document}